\documentclass{article}

\usepackage{arxiv}

\usepackage[utf8]{inputenc}
\usepackage[T1]{fontenc}
\usepackage{microtype}
\usepackage{amsmath}
\usepackage{amssymb}
\usepackage{nicefrac}
\usepackage{wasysym}
\usepackage{booktabs}
\usepackage{array}
\usepackage{tabularx}
\usepackage{xltabular}
\usepackage{multirow}
\usepackage{makecell}
\usepackage{siunitx}
\usepackage{adjustbox}
\usepackage{graphicx}
\usepackage{float}
\usepackage[table]{xcolor}
\usepackage{subfig}

\usepackage[numbers,square,sort&compress]{natbib}

\usepackage{hyperref}
\usepackage{doi}

\newcommand{\authorlink}[2]{%
    \begingroup
    \hypersetup{pdfborder={0 0 0},pdfborderstyle={}}%
    \href{#1}{#2}%
    \endgroup
}

\title{QRNet: QR-Specific Representation Learning for Direct Image-Based Quishing Detection}

\author{
    \authorlink{https://orcid.org/0009-0003-6646-6444}
    {\includegraphics[scale=0.06]{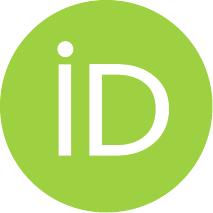}\hspace{1mm}John Harb} \\
    {\small Department of Electrical and} \\
    {\small Computer Engineering} \\
    {\small American University of Beirut} \\
    {\small Beirut, Lebanon} \\
    {\small\texttt{jah30@mail.aub.edu}}
    \And
    \authorlink{https://orcid.org/0000-0003-2241-8195}
    {\includegraphics[scale=0.06]{orcid.pdf}\hspace{1mm}Fouad Trad} \\
    {\small Department of Electrical and} \\
    {\small Computer Engineering} \\
    {\small Lebanese American University} \\
    {\small Byblos, Lebanon} \\
    {\small\texttt{fouad.trad@lau.edu.lb}}
    \And
    \authorlink{https://orcid.org/0000-0002-1939-2740}
    {\includegraphics[scale=0.06]{orcid.pdf}\hspace{1mm}Ali Chehab} \\
    {\small Department of Electrical and} \\
    {\small Computer Engineering} \\
    {\small American University of Beirut} \\
    {\small Beirut, Lebanon} \\
    {\small\texttt{chehab@aub.edu.lb}}
}

\date{}

\renewcommand{\shorttitle}{QRNet: QR-Specific Representation Learning for Direct Image-Based Quishing Detection}

\renewcommand{\undertitle}{}
\hypersetup{
pdftitle={QRNet: QR-Specific Representation Learning for Direct Image-Based Quishing Detection},
pdfsubject={QR Code Phishing Detection},
pdfauthor={John Harb, Fouad Trad, Ali Chehab},
pdfkeywords={Quishing, Deep Learning, Machine Learning, QR Code Phishing Detection, HOG Features},
}

\begin{document}
\maketitle

\begin{abstract}

QR codes are widely used for payments, authentication, advertising and information sharing, yet their visual form hides destination links and makes phishing attacks difficult to recognize. This study presents QRNet, a novel image-based deep learning architecture for detecting QR code phishing attacks or quishing, without decoding the embedded URL. QRNet was trained and evaluated on 1,005,738 QR images resized to ($224 \times 224$) pixels. QRNet achieved, on the test set, 90.53\% accuracy, 93.23\% precision, 89.98\% recall, a 91.58\% F1-score, a 90.37\% macro-F1 score and a 96.81\% ROC-AUC. QRNet numerically and statistically outperformed all twelve evaluated machine learning and deep learning baselines, including XGBoost, LightGBM, ResNet50, DenseNet121, EfficientNetB0 and a YOLO-based classifier. Its improvement over the strongest baseline was also highly significant according to DeLong test ($p=5.36\times10^{-47}$) and McNemar test ($p=3.66\times10^{-30}$). Robustness was evaluated under 27 distorted conditions involving blur, noise, JPEG compression, brightness and contrast changes, resizing, downscaling and occlusion, where QRNet remained stable under most distortions. Structural transfer experiments achieved macro-F1 scores of 97.56\% for mask pattern detection and 97.62\% for QR version detection, indicating that the learned representation captures meaningful QR specific structural information and can support tasks beyond phishing detection. For explainability analysis, QRNet decisions were examined using Grad-CAM, occlusion sensitivity, Integrated Gradients, LIME, local SHAP and global SHAP. Finally, QRNet was evaluated on an independent external dataset, showing strong zero-shot performance and further improvement after supervised fine-tuning, highlighting its ability to generalize and to adapt to unseen QR code images.
	
\end{abstract}


\section{Introduction}

QR codes are now widely used to connect users with digital services. They are used for payments, authentication, tickets, applications' downloads and access to online content \cite{309528,10.1145/3772318.3793213}. QR codes are easy to use because users can scan them instead of manually typing a URL or other information. However, this convenience also creates security risks. Unlike a normal link, the information stored in a QR code cannot be read directly by the user before scanning it. This makes it difficult to know whether a QR code is safe or malicious, just by its appearance \cite{309528,10.1145/3772318.3793213}.

The threat of QR code phishing has grown rapidly. According to Microsoft
Threat Intelligence, QR code phishing attacks increased from 7.6 million
in January 2026 to 18.7 million in March 2026, representing a 146\%
increase \cite{microsoft2026emailthreat}. Recent ESET data also show
that the threat remained high, with record levels of QR code phishing
reported during the first half of 2026, averaging around 100,000
detections per month and reaching the highest level in April
\cite{eset2026threatreport}. QR code phishing, commonly known as
quishing, can lead to credential theft, malware downloads, financial
fraud and other security problems \cite{Trad2027}. The problem is made
more difficult by the limited ability of users to recognize these
attacks. A recent study involving 1,876 participants found that only
13\% recognized fraudulent QR payment requests, compared with 46\%
when the same information was presented in a human readable form
\cite{309528}. These findings highlight the growing need for automatic
methods that can detect malicious QR codes before users interact with
their content.

To address these threats, phishing detection has traditionally focused
on URLs and web pages, and recent studies continue to use the URL itself
as a main feature for detection. These approaches analyze lexical
URL features, domain information, webpage content or learned
representations using machine learning and deep learning
\cite{nayak2025enhancing,11177149,JIANG2025111648,Chaudhuri2026}.
In quishing detection, a similar approach is often used after decoding
the QR code, where the extracted URL is analyzed using lexical features,
domain and host information, blacklist checks and webpage content,
often with machine learning/deep learning models
\cite{rafsanjani2023qsecr,MultiPhishNet2025,Almousa2026,11605248}.
At the same time, another direction has been explored that analyzes the QR
code itself without depending on its decoded content. Recent studies
have used pixel information, QR specific structural features and deep
visual models to distinguish benign and malicious QR codes
\cite{Trad2027,akram2025qrispreemptivenovelmethod,Alsulami2025,11621588,ozuberk2026malicious}. Existing work shows that the visual structure of QR codes can provide useful information for quishing detection.

Despite this progress, several limitations remain. Many existing studies
use URL decoding, which requires the QR code content to be extracted
before detection and does not directly analyze the visual structure of
the QR code image. Other studies use direct QR code analysis, but many
rely on manually selected features, traditional machine learning or commonly used deep learning models. Among the studies reviewed in this work, none proposed a deep learning architecture designed specifically for direct quishing detection from QR code images. In addition, most studies use relatively small QR image datasets, which can limit the evaluation of how well their methods generalize to different QR codes. Robustness to image distortions, statistical validation, explainability and evaluation on independent external datasets are also limited in most of
the existing work. None of the reviewed studies examined transfer to
other QR specific tasks. Among the studies reviewed in this work, none
combines large scale image only detection, a broad machine learning and
deep learning comparison, statistical validation, robustness analysis,
explainability, external evaluation and QR specific transfer within a
single framework. 

To address these gaps, this study proposes QRNet, a novel deep learning
architecture designed specifically for QR code image analysis and
quishing attack detection. QRNet learns directly from QR code images without decoding the embedded content or requiring manually selected features. The model is trained and evaluated on more than one million QR code images and compared with 12 machine learning and deep learning baseline models. Statistical tests are used to determine whether the differences between QRNet and the strongest baseline are significant. Robustness analysis examines the model under different image distortions, while explainability methods provide insight into the regions and learned
features that influence its predictions. QRNet is also evaluated on an
independent external dataset to examine its ability to generalize to
unseen QR code images. In addition, its learned features are evaluated
on QR version and mask pattern classification to study whether they can
support other QR specific tasks.

The main contributions of this study are as follows:

\begin{itemize}

    \item To the best of our knowledge, we propose the first QR specific
    deep learning architecture for direct benign/malicious QR code image
    classification.

    \item We provide the largest image-based quishing benchmark reported
    to date, using more than one million QR code images and 13 machine
    learning and deep learning models.

    \item We evaluate QRNet using statistical tests, robustness analysis
    and multiple explainability methods.

    \item We study the transfer of QRNet to QR version and mask pattern classification.

    \item We evaluate QRNet on an independent external QR code dataset
    using zero-shot testing and fine-tuning.

\end{itemize}

The remainder of this paper is organized as follows.
Section \ref{sec:related-work} reviews related work on QR code phishing
detection and discusses the main gaps in existing studies.
Section \ref{sec:methodology} presents the proposed QRNet architecture,
dataset preparation, preprocessing and experimental setup.
Section \ref{sec:results} presents and discusses the main classification
results, statistical analysis, robustness evaluation, explainability,
QR specific transfer tasks and external dataset evaluation.
Finally, Section \ref{sec:conclusion} concludes the paper and presents
directions for future work.

\section{Related Work}
\label{sec:related-work}

Research on QR code phishing (quishing) detection has evolved along several distinct directions. Many studies focus on analyzing URLs, domain characteristics, and webpage content, typically after extracting and decoding the information embedded within QR codes. In contrast, other studies investigate the visual and structural properties of QR code images directly, without relying on content decoding. More recent approaches increasingly adopt multimodal strategies that integrate heterogeneous sources of information, including QR code images, decoded URLs, optical character recognition (OCR) text, payment-related attributes, contextual or location information, and predefined security rules. A further line of research addresses application-oriented defenses, such as secure QR code scanners, authenticity and integrity verification mechanisms, and broader phishing prevention frameworks. Reflecting these methodological differences, this section organizes the existing literature into four main categories: URL-based approaches, direct image-based approaches, multimodal approaches, and application-specific defense mechanisms.

\subsection{URL and Decoded QR-Based Phishing Detection}

URL-based phishing detection uses information such as URL structure, domain properties, host information and webpage content. The authors in \cite{nayak2025enhancing} applied feature selection before comparing FNN, DNN, Wide/Deep and TabNet models and tested the selected model on a second URL dataset. However, the method relies on handcrafted URL and domain information and does not examine QR images or their visual structure. Similarly, the researchers in \cite{kumar2025hybridstack} combined several machine learning models with an MLP meta classifier using handcrafted URL, domain and webpage features. Although some URLs were extracted from QR codes, they were merged with directly collected URLs and the final detector remained dependent on decoded content rather than QR image patterns. This study \cite{abir2026bert} compared traditional machine learning models with a fine-tuned BERT model for multiclass malicious URL detection and used LIME to explain individual predictions. The study used 130,080 URLs, while our QRNet dataset contains 1,005,738 QR images, making it almost eight times larger. Moreover, the study did not include QR processing, QR image robustness testing, external dataset evaluation, or statistical comparison. PhishGuard \cite{savitha2026phishguard} used an RBF-kernel SVM with URL and webpage content features, including HTML structure, JavaScript behavior, form actions and iframe usage. However, it did not provide QR specific evaluation or compare machine learning and deep learning models. Finally, the authors of \cite{Kusuma2026} compared Random Forest, Decision Tree and Naïve Bayes in a simulated QRIS quishing setting. The models used a general phishing URL dataset whose samples were assumed to originate from QR scans, rather than real QRIS transactions or an implemented QR scanning system. These studies show the usefulness of URL-based classification, but their decisions remain based on URL or webpage information and do not directly capture the visual and structural properties of QR codes. 

Other studies place URL-based detection inside QR scanning or QR processing systems. QsecR \cite{rafsanjani2023qsecr} is an Android scanner that decodes QR codes and examines the extracted URLs using redirection checks, blacklists, handcrafted features and external services. Its fixed rules and external dependencies may limit its ability to adapt to new attacks. MultiPhishNet \cite{MultiPhishNet2025} combines word and character level URL embeddings with multilingual HTML embeddings using CNNs and multi-head attention. However, it requires successful QR decoding and live webpage access and does not provide QR image robustness testing, external dataset evaluation, statistical validation or dedicated explainability. The incremental framework in \cite{10.1145/3802927.3802959} uses BERT, CNN and BiLSTM components to classify decoded URLs and updates the upper model layers using recent phishing data. Although it was evaluated using a separate QR dataset with rotation, blur and noise, the classifier itself remained URL-based and the adaptation experiment covered only one update without formal statistical or explainability analysis. The technique in \cite{Almousa2026} also scans QR codes before classifying their URLs using BERT-based representations and Logistic Regression. It periodically retrains the model with URLhaus records, but it explicitly excludes physical and visual QR manipulation and does not evaluate QR image distortions or provide statistical and explainability analyses. Likewise, the browser-based QR and NFC platform in \cite{al2026feature} extracts the embedded URL, computes 30 handcrafted features and compares several classifiers before selecting XGBoost. Despite its practical scanning interface, the trained detector remains dependent on decoded URL features and does not evaluate visual distortions, external transfer or dedicated explainability. So, these systems support QR related use cases but still require successful content extraction. QRNet fills these gaps by detecting malicious QR codes directly from their images, without decoding URLs or accessing webpages. It also tests the model under image distortions, evaluates it on another dataset, uses statistical tests and includes several explainability methods.

\subsection{Direct Image-Based Detection of Malicious QR Codes}  

Several studies have examined malicious QR codes through their visual or structural patterns without relying only on decoded URLs. The authors in \cite{akram2025qrispreemptivenovelmethod} proposed QRiS, which converts QR images into binary module grids and extracts 24 structural and statistical features for classification using XGBoost and Random Forest. Although the study used 400,000 synthetic QR codes, the method depends on a fixed set of handcrafted features and supports only black and white QR codes with square modules. This may limit its use with colored, stylized, or visually modified QR codes. Its evaluation is also restricted to two machine learning models, without deep learning comparisons or external validation. The study in \cite{Alsulami2025} combines AlexNet, PCA and recurrent networks to classify 21,000 QR images and also presents a Streamlit prototype. However, the pipeline reduces the AlexNet features to 50 fixed components before classification, rather than learning the complete image representation end-to-end. Moreover, the two datasets contain only QR versions 1 and 2, and the study does not examine the robustness against distortion, explainability, or structural generalization. The research in \cite{tashtoush2025comparing} compares decoded URL models with MobileNetV3Large and EfficientNetB0 using the same 1,000 QR images. The image-based models avoid URL decoding and provide faster predictions. Despite these advantages, the very small dataset limits the strength of the comparison, while the image transformations used during training are not evaluated as a separate robustness experiment. The study in \cite{Trad2027} analyzes 9,987 QR codes by treating each pixel as an input feature and comparing several traditional ML models. The authors also map feature importance onto the QR grid and remove non informative pixels. However, all QR codes use version 13 and were generated in the same fixed format, which limits their representation of real world QR code variations. The study also excludes deep learning models and does not evaluate robustness to image distortions or transfer to other QR structural tasks. The research in \cite{maree2026classification} trains a custom CNN using 632,508 balanced QR images. Although the dataset is large, the images are generated under controlled conditions and the evaluation is limited to one CNN architecture. Therefore, the study does not show whether its results remain stable across different model families, image distortions and it provides no dedicated explainability. The study in \cite{ozuberk2026malicious} offers a broader evaluation by comparing several ML and deep learning models on 1,000 QR images and includes model explainability. Yet, the dataset contains mostly standard QR codes and the study does not investigate performance under noisy, low resolution or real-time scanning conditions. Moreover, external validation, formal statistical testing and QR structural generalization are also absent. Taken together, these studies show the value of direct image-based detection, but they remain limited by handcrafted or fixed representations, restricted QR settings, controlled or small datasets, narrow model evaluation and limited evidence of robustness and generalization. QRNet addresses these gaps through large-scale representation learning, multiple ML and deep learning baselines, systematic distortion testing, explainability, formal statistical validation, external adaptation and separate QR structural generalization tasks.

\subsection{Multimodal and Application Specific QR Threat Detection}

Recent works extend QR threat detection by combining visual,
payload, contextual or behavioral information, while others address
specific payment and scanning scenarios. The framework in
\cite{Naik2025} combines features from decrypted QR content with
ResNet50-based image analysis and fuses Gradient Boosting and deep
learning outputs for multi-class threat detection. However, the paper
does not report the total dataset size or class distribution, and the
presented confusion matrix does not clearly represent all seven attack
categories described in the method, which makes the full multi-class
evaluation difficult to assess. AP3X \cite{rivas2025securing}
accepts uploaded QR images, camera captures or direct URLs, extracts
the embedded link and applies several ML models using
24 URL features, supported by VirusTotal and urlscan.io. Although this
provides a broader security check, the classifiers mainly depend on
decoded URL information and external services rather than direct
analysis of a QR image. The approach in \cite{Supriya2025} follows a
sequential pipeline in which YOLOv8 locates the code, DenseNet121
classifies it as genuine or counterfeit and a separate module checks
the extracted URL for phishing. This combines image and URL analysis,
but errors in the earlier stages can affect the final result, and the
phishing decision still requires URL decoding. In
\cite{saputra2026deteksi}, the researchers developed a QRIS specific
fraud detection system that combines YOLOv11, PaddleOCR, decoded EMVCo
information, layout rules and GPS verification. Its practical focus is
useful, but the dataset contains only 898 QRIS images, including
manually created anomalies, while the complete system is evaluated on
only 130 samples and is closely tied to the Indonesian Static QRIS MPM
format. The multimodal design in \cite{shashank2026quishing} combines
QR images, decoded URL features and UPI transaction metadata through
late fusion, yet the reported experiments mainly cover the QR branch
using 9,987 fixed version 13 QR codes, without separate results for the
URL, transaction or final fused components. WHALE in
\cite{s26144412} instead analyzes the QR decoded URL inside an isolated
Sandbox WebView by combining static URL, HTML and DOM features with
dynamic credential flow events. While this supports practical on device protection, it does not model QR image structure, evaluates its static and dynamic stages separately, and limits the dynamic dataset to credential request flows with observable submission of test credentials. Taken together, these approaches improve QR security through
multimodal or application specific designs, but they remain limited by
decoded URL dependence, sequential processing, narrow datasets,
incomplete end-to-end evaluation or specialized deployment settings.

\subsection{Synthesis and Research Gap}

The reviewed studies show that QR quishing detection still has several
important limitations. Many studies rely on decoded URLs, webpage
content, transaction information or external security services rather
than detecting threats directly from QR images. Existing image-based
methods often use small, generated or
fixed version datasets that may not reflect the diversity of real QR
codes. They also commonly compare only a few models, use limited visual
representations, and provide little evidence of robustness to image
distortions or transfer to QR codes from a different source. Some
studies examine only ML or deep learning, which
limits comparison between the two model families. Other works provide
limited explainability, do not apply statistical tests or report
results only on their original test set. Multimodal and sequential
systems use more information, but errors in one stage may affect
later stages and the complete system is not always evaluated end-to-end. None of the reviewed works combines large scale image only
detection, broad ML and DL comparison, formal statistical validation,
explainability, realistic distortion testing, external adaptation and
QR structural evaluation within one framework. QRNet addresses these
gaps through a QR specific deep learning model that detects quishing
directly from QR images without decoding their embedded content. It is
trained and evaluated on more than one million QR images and
 numerically outperformed the 12 ML and deep learning
baselines evaluated. DeLong and McNemar tests further confirmed that
its improvement over the strongest evaluated baseline was also statistically significant. The evaluation also includes multiple explainability methods, 27 realistic QR image distortion conditions, and adaptation to an external QR dataset through fine-tuning. Moreover, the learned representation is successfully applied to QR version and mask pattern classification, showing that QRNet can support several QR specific tasks beyond quishing detection.

To summarize the reviewed studies and make the comparison clearer,
Table \ref{tab:related-work-comparison} presents a comparative overview
of the selected literature. It compares the detection tasks, dataset size and type, model families and evaluation methods, while highlighting the main gaps addressed by QRNet.

\begin{table}[p]
\caption{Comparative Analysis of Related Work on Phishing and Quishing Detection.}
\label{tab:related-work-comparison}
\centering
\begingroup

\fontsize{8}{9.6}\selectfont
\setlength{\tabcolsep}{1.0pt}
\renewcommand{\arraystretch}{1.05}

\rowcolors{2}{gray!10}{white}

\newcommand{\QRHdr}[1]{\makebox[\linewidth][c]{\textbf{#1}}}

\begin{tabularx}{\textwidth}{@{}
  >{\raggedright\arraybackslash}p{0.95cm}
  >{\centering\arraybackslash}p{0.70cm}
  @{\hspace{5pt}}
  >{\raggedright\arraybackslash}X
  >{\centering\arraybackslash}p{0.80cm}
  @{\hspace{4pt}}
  >{\raggedright\arraybackslash}X
  >{\centering\arraybackslash}p{0.70cm}
  @{\hspace{4pt}}
  >{\raggedright\arraybackslash}X
  @{\hspace{3pt}}
  >{\centering\arraybackslash}p{1.20cm}
  @{\hspace{2pt}}
  >{\centering\arraybackslash}p{0.60cm}
  @{\hspace{2pt}}
  >{\centering\arraybackslash}p{0.60cm}
  @{\hspace{2pt}}
  >{\centering\arraybackslash}p{0.70cm}
  @{\hspace{2pt}}
  >{\centering\arraybackslash}p{0.80cm}
  @{\hspace{2pt}}
  >{\centering\arraybackslash}p{0.75cm}@{}}

\hline\noalign{\smallskip}

\rowcolor{white}
\QRHdr{Study} & \QRHdr{Year} & \QRHdr{Input Type} &
\QRHdr{QSM} & \QRHdr{Task} & \QRHdr{IOD} &
\QRHdr{DS} & \QRHdr{ML/DL} & \QRHdr{SV} &
\QRHdr{DR} & \QRHdr{XAI} & \QRHdr{EDE} & \QRHdr{QST} \\

\noalign{\smallskip}\hline\noalign{\smallskip}

\cite{rafsanjani2023qsecr} & 2023 & Decoded QR URL & $\LEFTcircle$ &
Malicious URL detection & $\times$ & 5,500 URLs &
$\times/\times$ & $\times$ & $\times$ & $\times$ & $\LEFTcircle$ &
$\times$ \\[2pt]

\cite{nayak2025enhancing} & 2025 & URL & $\times$ &
Phishing URL detection & $\times$ & 58,645 URLs &
$\times/\checkmark$ & $\times$ & $\times$ & $\LEFTcircle$ &
$\checkmark$ & $\times$ \\[2pt]

\cite{11177149} & 2025 & URL + HTML features & $\times$ &
Phishing URL detection & $\times$ & 101,063 URLs &
$\checkmark/\times$ & $\checkmark$ & $\times$ & $\LEFTcircle$ &
$\times$ & $\times$ \\[2pt]

\cite{JIANG2025111648} & 2025 & URL + HTML/JavaScript & $\times$ &
Phishing webpage detection & $\times$ & 70,557 URL--HTML webpages &
$\LEFTcircle/\checkmark$ & $\times$ & $\times$ & $\times$ & $\checkmark$ &
$\times$ \\[2pt]

\cite{MultiPhishNet2025} & 2025 & Decoded URL-HTML & $\LEFTcircle$ &
Phishing website detection & $\times$ & 100,046 URL--HTML &
$\times/\checkmark$ & $\times$ & $\times$ & $\times$ & $\times$ &
$\times$ \\[2pt]

\cite{kumar2025hybridstack} & 2025 & URL & $\times$ &
Malicious URL detection & $\times$ & 14,013 URLs &
$\checkmark/\LEFTcircle$ & $\times$ & $\times$ & $\times$ & $\times$ &
$\times$ \\[2pt]

\cite{11264995} & 2025 & QR images / Decoded QR URLs & $\checkmark$ &
QR validity and malicious URL detection & $\LEFTcircle$ &
300,000 QR images + 446,176 URLs & $\times/\checkmark$ &
$\times$ & $\times$ & $\times$ & $\LEFTcircle$ & $\times$ \\[2pt]

\cite{akram2025qrispreemptivenovelmethod} & 2025 & 24 QR structural features & $\checkmark$ &
Quishing detection & $\checkmark$ & 2 $\times$ 200,000 QR code images &
$\checkmark/\times$ & $\times$ & $\times$ & $\LEFTcircle$ & $\times$ &
$\times$ \\[2pt]

\cite{Alsulami2025} & 2025 & 50 QR features & $\checkmark$ &
Quishing detection & $\checkmark$ & 21,000 QR code images &
$\checkmark/\checkmark$ & $\times$ & $\times$ & $\times$ & $\times$ &
$\times$ \\[2pt]

\cite{tashtoush2025comparing} & 2025 & QR image / Decoded QR URL & $\checkmark$ &
Quishing detection & $\LEFTcircle$ & 1,000 QR code images &
$\checkmark/\checkmark$ & $\times$ & $\times$ & $\times$ & $\times$ &
$\times$ \\[2pt]

\cite{Naik2025} & 2025 & QR payload features / QR images & $\checkmark$ &
Quishing detection & $\LEFTcircle$ & Not reported &
$\LEFTcircle/\LEFTcircle$ & $\times$ & $\times$ & $\times$ & $\times$ &
$\times$ \\[2pt]

\cite{rivas2025securing} & 2025 & 24 decoded QR URL features & $\LEFTcircle$ &
Malicious URL detection & $\times$ & $>600{,}000$ URLs + 1,000 QR code images &
$\checkmark/\LEFTcircle$ & $\times$ & $\times$ & $\times$ &
$\checkmark$ & $\times$ \\[2pt]

\cite{Supriya2025} & 2025 & QR images / Decoded QR URL & $\checkmark$ &
QR authenticity and phishing detection & $\LEFTcircle$ & 200,000 QR code images &
$\times/\LEFTcircle$ & $\times$ & $\times$ & $\times$ &
$\times$ & $\times$ \\[2pt]

\cite{saputra2026deteksi} & 2026 & QRIS image / decoded QR data & $\checkmark$ &
QRIS fraud detection & $\LEFTcircle$ & 898 QRIS images &
$\times/\checkmark$ & $\times$ & $\times$ & $\times$ & $\times$ &
$\times$ \\[2pt]

\cite{shashank2026quishing} & 2026 & QR / URL / UPI features & $\checkmark$ &
UPI quishing detection & $\LEFTcircle$ & 9,987 QR codes; others NR &
$\checkmark/\times$ & $\times$ & $\times$ & $\times$ &
$\times$ & $\times$ \\[2pt]

\cite{s26144412} & 2026 & Decoded QR URL / page events & $\LEFTcircle$ &
Phishing URL detection & $\times$ & 11,480 webpages + 2,837 flows &
$\checkmark/\checkmark$ & $\times$ & $\LEFTcircle$ &
$\LEFTcircle$ & $\times$ & $\times$ \\[2pt]

\cite{Trad2027} & 2026 & Flattened QR pixels & $\checkmark$ &
Quishing detection & $\checkmark$ & 9,987 QR code images &
$\checkmark/\times$ & $\times$ & $\times$ & $\LEFTcircle$ & $\times$ &
$\times$ \\[2pt]

\cite{maree2026classification} & 2026 & QR code images & $\checkmark$ &
Quishing detection & $\checkmark$ & 632,508 QR code images &
$\times/\LEFTcircle$ & $\times$ & $\times$ & $\times$ & $\times$ &
$\times$ \\[2pt]

\cite{ozuberk2026malicious} & 2026 & QR image features / QR images & $\checkmark$ &
Quishing detection & $\checkmark$ & 1,000 QR code images &
$\checkmark/\checkmark$ & $\times$ & $\times$ & $\checkmark$ & $\times$ &
$\times$ \\[2pt]

\cite{10.1145/3802927.3802959} & 2026 & Decoded QR URL & $\LEFTcircle$ &
Phishing URL detection & $\times$ & 220,000 URLs + 10,001 QR code images &
$\times/\checkmark$ & $\times$ & $\checkmark$ & $\times$ & $\checkmark$ &
$\times$ \\[2pt]

\cite{abir2026bert} & 2026 & URL & $\times$ &
4 class URL classification & $\times$ & 130,080 URLs &
$\checkmark/\checkmark$ & $\times$ & $\times$ & $\checkmark$ & $\times$ &
$\times$ \\[2pt]

\cite{Almousa2026} & 2026 & Decoded QR URL & $\LEFTcircle$ &
Malicious URL detection & $\times$ & 25,428 URLs &
$\LEFTcircle/\LEFTcircle$ & $\times$ & $\times$ & $\times$ & $\checkmark$ &
$\times$ \\[2pt]

\cite{savitha2026phishguard} & 2026 & URL + webpage content & $\times$ &
Phishing website detection & $\times$ & Not clearly reported &
$\LEFTcircle/\times$ & $\times$ & $\times$ & $\times$ & $\times$ &
$\times$ \\[2pt]

\cite{al2026feature} & 2026 & Decoded QR URL & $\LEFTcircle$ &
Malicious URL detection & $\times$ & 20,000 URLs &
$\checkmark/\checkmark$ & $\times$ & $\times$ & $\times$ & $\times$ &
$\times$ \\[2pt]

\cite{11605248} & 2026 & Decoded QR URL & $\LEFTcircle$ &
Malicious URL detection & $\times$ & 336,749 URLs &
$\LEFTcircle/\times$ & $\times$ & $\LEFTcircle$ & $\LEFTcircle$ & $\times$ &
$\times$ \\[2pt]

\cite{11621588} & 2026 & QR code images & $\checkmark$ &
Quishing detection & $\checkmark$ & 632,508 QR code images &
$\times/\LEFTcircle$ & $\times$ & $\times$ & $\times$ & $\times$ &
$\times$ \\[2pt]

\cite{Chaudhuri2026} & 2026 & URL & $\times$ &
Phishing URL detection & $\times$ & 650,000 URLs &
$\checkmark/\times$ & $\checkmark$ & $\checkmark$ & $\LEFTcircle$ &
$\times$ & $\times$ \\[2pt]

\cite{Kusuma2026} & 2026 & URL & $\times$ &
Phishing URL detection & $\times$ & 88,647 URLs &
$\checkmark/\times$ & $\times$ & $\times$ & $\LEFTcircle$ & $\times$ &
$\times$ \\[2pt]

\noalign{\smallskip}\hline\noalign{\smallskip}

\textbf{QRNet} & \textbf{2026} & \textbf{QR code image} &
$\boldsymbol{\checkmark}$ &
\textbf{Quishing detection} & $\boldsymbol{\checkmark}$ &
\textbf{1,005,738 QR code images} &
$\boldsymbol{\checkmark/\checkmark}$ &
$\boldsymbol{\checkmark}$ &
$\boldsymbol{\checkmark}$ &
$\boldsymbol{\checkmark}$ &
$\boldsymbol{\checkmark}$ &
$\boldsymbol{\checkmark}$ \\

\noalign{\smallskip}\hline

\end{tabularx}

\par\vspace{4pt}
\noindent
\begin{minipage}{\textwidth}
\fontsize{8}{9.6}\selectfont
Input: input type;
Task: detection task;
IOD: image only detection without URL decoding;
DS: dataset size;
QSM: QR specific model;
ML/DL: machine learning/deep learning comparison;
SV: statistical validation;
DR: distortion robustness;
XAI: explainability;
EDE: external dataset evaluation;
QST: QR structural transfer.
$\checkmark$: fully addressed;
$\LEFTcircle$: partially addressed;
$\times$: not addressed.
\end{minipage}

\endgroup
\end{table}

\section{Methodology}
\label{sec:methodology}

\subsection{Dataset}

The dataset used in this study is CIC-Trap4Phish
\cite{nejati2026cictrap4phishunifiedmultiformatdataset}, a large scale
dataset developed for phishing and quishing attack detection. It
consists of 1,005,738 labeled QR code images, including 429,976 benign
and 575,762 malicious samples. The QR images were generated from benign
and malicious URLs collected from multiple reliable public sources,
providing a diverse collection of encoded URL patterns. The dataset
was selected because it is the largest and most recent dataset available
for QR-based quishing research, providing sufficient scale and diversity
for training and evaluating quishing attack detection using ML and DL models.

\begin{figure}[!htbp]
\centering

\subfloat[Malicious QR images\label{fig:eda-malicious}]{%
    \includegraphics[width=0.48\textwidth]{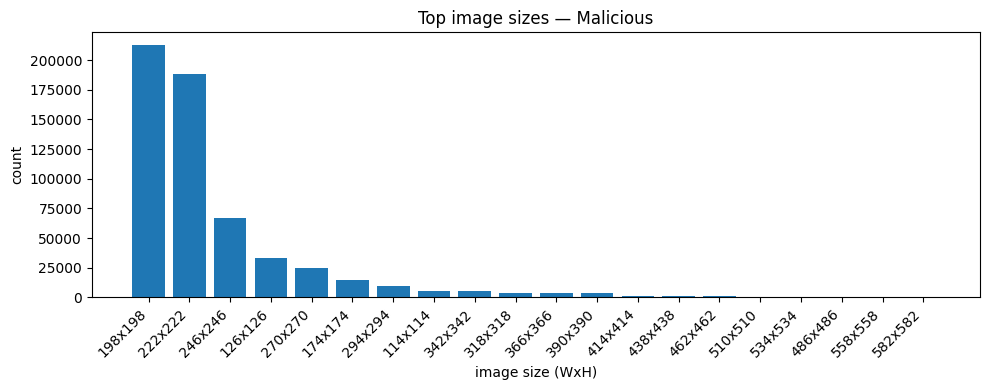}%
}
\hfill
\subfloat[Benign QR images\label{fig:eda-benign}]{%
    \includegraphics[width=0.48\textwidth]{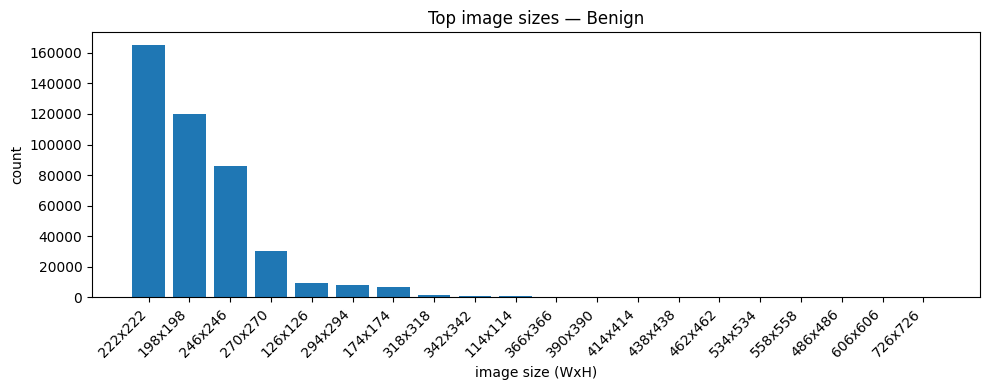}%
}

\caption{Distribution of Raw QR Image Dimensions by Class Before Preprocessing.}
\label{fig:eda-image-sizes}

\end{figure}

Before preprocessing, we performed a preliminary exploratory analysis
of all 1,005,738 QR images to examine the distribution of their original
sizes. As shown in Figure \ref{fig:eda-malicious} and
Figure \ref{fig:eda-benign}, most images in both classes are concentrated
in a small number of square resolutions, while the remaining sizes occur
much less frequently. In the malicious class, the most common resolution
is ($198 \times 198$), followed by ($222 \times 222$) and
($246 \times 246$). In the benign class, ($222 \times 222$) is the most
common resolution, followed by ($198 \times 198$) and
($246 \times 246$). Other sizes, including ($126 \times 126$),
($270 \times 270$), and ($294 \times 294$), are also present but appear
less often in both classes.

For all ML, DL and QRNet models, the
dataset was divided using a stratified 70\%/15\%/15\%
train/val/test split with a fixed random seed of 42. This
preserved the benign and malicious class proportions in each set and
ensured that all models were evaluated using the same data partitions.

\subsection{Preprocessing and HOG Features}

\subsubsection{Image Resizing}

\begin{figure}[!htbp]
    \centering

    \subfloat[Resized malicious QR code.
    \label{fig:resized-malicious}]{%
        \includegraphics[width=0.40\textwidth]{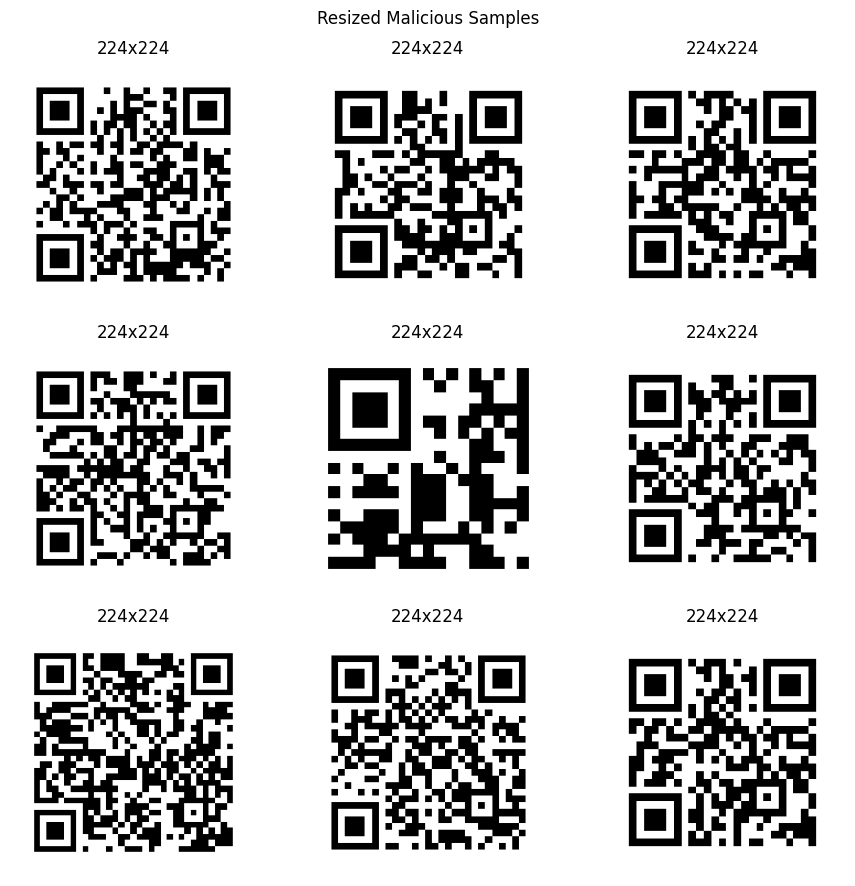}%
    }
    \hspace{0.08\textwidth}
    \subfloat[Resized benign QR code.
    \label{fig:resized-benign}]{%
        \includegraphics[width=0.40\textwidth]{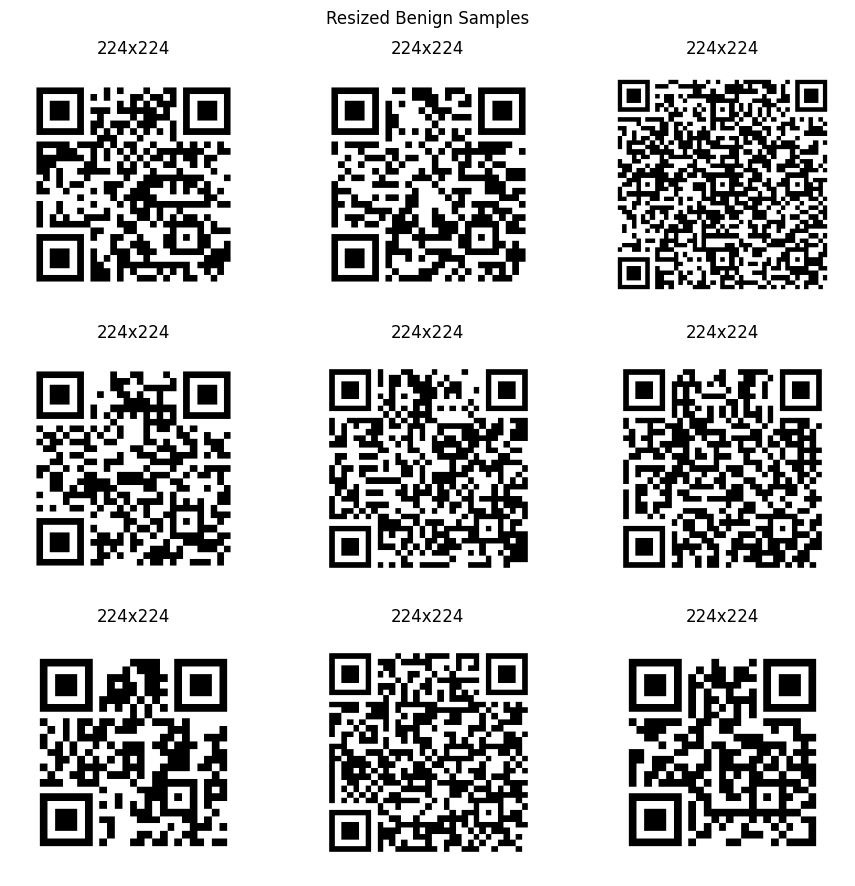}%
    }

    \caption{Examples of QR images resized to $224 \times 224$ pixels:
    (a) malicious samples and (b) benign samples.}
    \label{fig:resized-qr-examples}

\end{figure}

EDA analysis showed that the original QR images had different
dimensions. Therefore, all images were resized to ($224 \times 224$)
pixels before feature extraction and model training. This provided a
consistent input size for ML, DL and QRNet
models and prevented differences in the original image dimensions
from affecting the model comparison. Figure \ref{fig:resized-qr-examples}
shows examples of malicious and benign QR images after resizing.

\subsubsection{HOG Feature Extraction}

After resizing, the Histogram of Oriented Gradients (HOG) feature vectors
were extracted for the ML models. Each ($224 \times 224$) QR image was converted to grayscale and its pixel values were normalized to the range $[0,1]$. HOG was applied using nine orientation bins, cells of ($16 \times 16$) pixels, blocks of ($2 \times 2$) cells, and L2-Hys block normalization. For each grayscale image $I(x,y)$, the horizontal and vertical intensity changes were calculated as

\begin{equation}
G_x(x,y)=I(x+1,y)-I(x-1,y),
\qquad
G_y(x,y)=I(x,y+1)-I(x,y-1).
\end{equation}

The strength and direction of the intensity change at each pixel were
then obtained using:

\begin{equation}
M(x,y)=\sqrt{G_x(x,y)^2+G_y(x,y)^2},
\qquad
\theta(x,y)=\operatorname{atan2}\!\left(G_y(x,y),G_x(x,y)\right).
\end{equation}

Within each cell, the gradient directions were grouped into a
nine-bin histogram. Each pixel contributed to the histogram according
to its gradient magnitude:

\begin{equation}
h_b=
\sum_{(x,y)\in\mathrm{cell}}
M(x,y)\,w_b\!\left(\theta(x,y)\right),
\qquad b=1,\ldots,9,
\end{equation}

where $w_b(\cdot)$ is the weight assigned to orientation bin $b$.
The histograms from the four cells in each block were joined into a
vector $\mathbf{v}$ and normalized using L2-Hys:

\begin{equation}
\mathbf{v}_{\mathrm{norm}}
=
\frac{\mathbf{v}}
{\sqrt{\lVert\mathbf{v}\rVert_2^2+\epsilon^2}},
\qquad
\mathbf{v}_{\mathrm{Hys}}
=
\min\left(\mathbf{v}_{\mathrm{norm}},\tau\right),
\end{equation}

\begin{equation}
\mathbf{v}_{\mathrm{final}}
=
\frac{\mathbf{v}_{\mathrm{Hys}}}
{\sqrt{\lVert\mathbf{v}_{\mathrm{Hys}}\rVert_2^2+\epsilon^2}},
\qquad \tau=0.2.
\end{equation}

The final length of the feature vector was determined from the HOG grid.
Dividing the image size by the cell size produced:

\begin{equation}
\frac{224}{16}=14,
\end{equation}

giving ($14 \times 14$) cells. Since each block contained
($2 \times 2$) cells and adjacent blocks overlapped, the number of blocks
was:

\begin{equation}
(14-2+1)\times(14-2+1)=13\times13.
\end{equation}

Each block contained

\begin{equation}
2\times2\times9=36
\end{equation}

features. Therefore, each QR image was represented by:

\begin{equation}
13\times13\times36=6084
\end{equation}

HOG features per image. To process the full dataset efficiently, these features were stored in compressed chunks of 25,000 samples. The resulting
6,084 dimensional feature vectors were used only as inputs to the
ML models, while the DL models and QRNet
received the resized QR images directly.

\subsection{Models and Hyperparameter Settings}

\subsubsection{Machine Learning Models}

Eight ML models were evaluated using the
6,084 dimensional HOG feature vectors extracted from the QR code images
as input. These models included Logistic Regression, Support Vector
Machine, Random Forest, XGBoost, LightGBM, CatBoost, AdaBoost and
Naive Bayes. The selected models represent different learning
approaches, including linear classification, probabilistic
classification, bagging and boosting. Together, they provide a varied
set of ML baselines for distinguishing between benign and
malicious QR codes and for comparison with the DL models and QRNet.

\subsubsection{Deep Learning Models}
Four DL models were evaluated directly using the resized QR
code images as input. These models included EfficientNet-B0, ResNet50, DenseNet121 and YOLO. The selected models represent different
DL architectures and feature extraction approaches. They provide a varied set of deep learning baselines for comparing their
performance with the ML models and QRNet.

\subsubsection{Proposed QRNet Architecture}

\begin{figure}[!htbp]
    \centering
    \includegraphics[width=\textwidth]{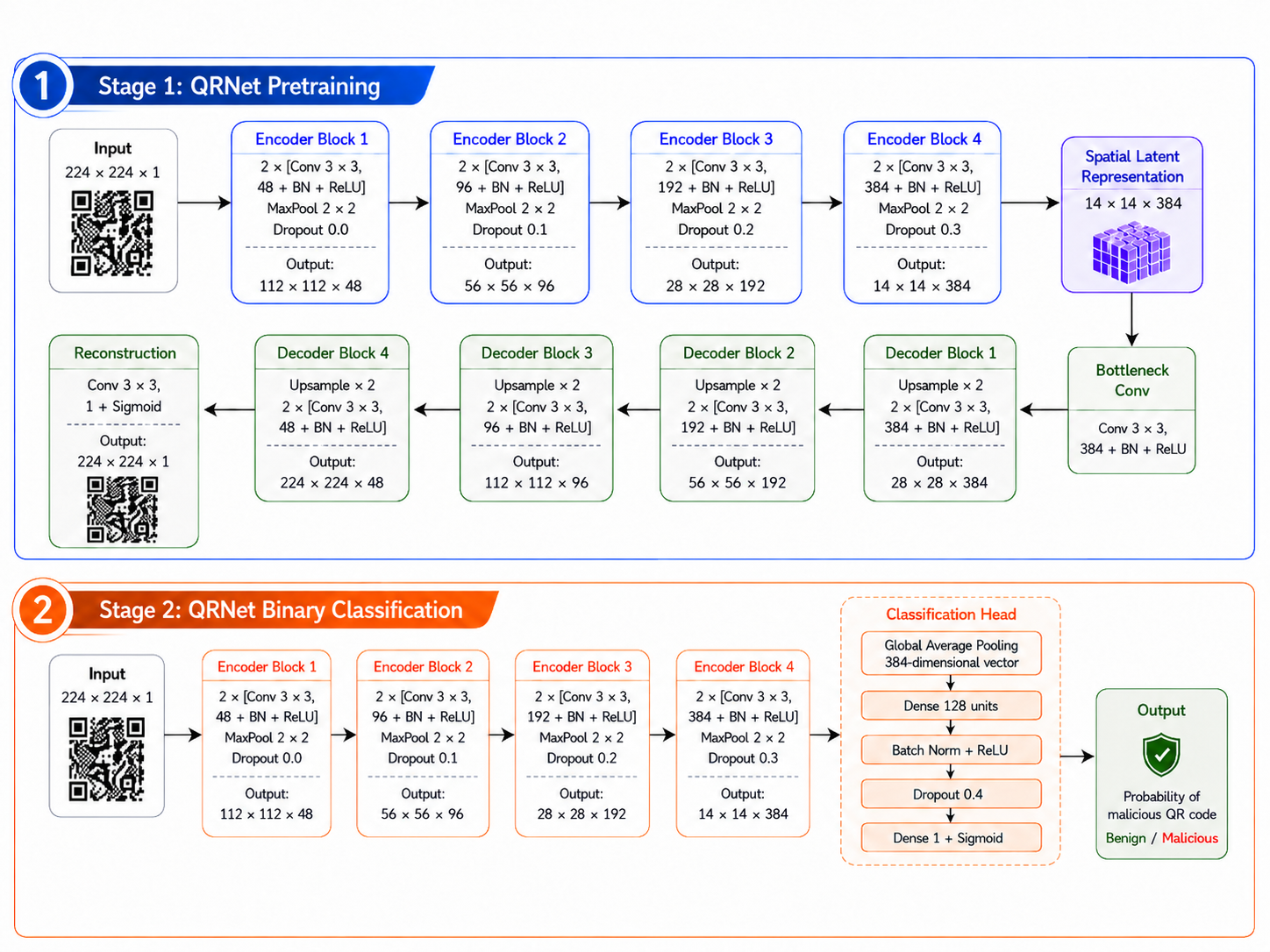}
    \caption{Architecture of the proposed QRNet model.}
    \label{fig:qrnet-architecture}
\end{figure}

QRNet is a custom convolutional neural network designed to learn
features directly from QR code images. Its encoder is first trained as
part of a convolutional autoencoder without using benign or malicious
labels. The pretrained encoder is then connected to a  binary classification head and adapted to distinguish between benign
and malicious QR codes. The complete architecture and its two main
phases are illustrated in Figure \ref{fig:qrnet-architecture}.

The model receives a grayscale QR code image of size
\(224 \times 224 \times 1\). The encoder contains four convolutional
blocks with \(48\), \(96\), \(192\) and \(384\) filters,
respectively. Each block contains two \(3 \times 3\) convolutional
layers, with each convolution followed by batch normalization and ReLU
activation. A \(2 \times 2\) max pooling layer is then applied to
reduce the spatial dimensions. Dropout rates of \(0\), \(0.1\),
\(0.2\) and \(0.3\) are used after the pooling operations of the four
blocks, respectively.

The spatial dimensions are progressively reduced as follows:
\[
224 \times 224
\rightarrow 112 \times 112
\rightarrow 56 \times 56
\rightarrow 28 \times 28
\rightarrow 14 \times 14.
\]
The final encoder output is therefore a spatial feature representation
of size
\[
14 \times 14 \times 384.
\]
During autoencoder pretraining, this representation is passed to a
decoder that reconstructs the original QR code image. The decoder first
applies a \(3 \times 3\) bottleneck convolution with \(384\) filters,
followed by batch normalization and ReLU activation. Then, it uses four
decoder blocks that progressively restore the spatial resolution. Each
decoder block performs two fold upsampling followed by two
convolution, batch normalization, and ReLU operations. The numbers of
filters used across the decoder blocks are \(384\), \(192\), \(96\)
and \(48\), respectively. The spatial dimensions are increased as
\[
14 \times 14
\rightarrow 28 \times 28
\rightarrow 56 \times 56
\rightarrow 112 \times 112
\rightarrow 224 \times 224.
\]

\begin{figure}[!htbp]
    \centering
    \includegraphics[width=0.90\textwidth]{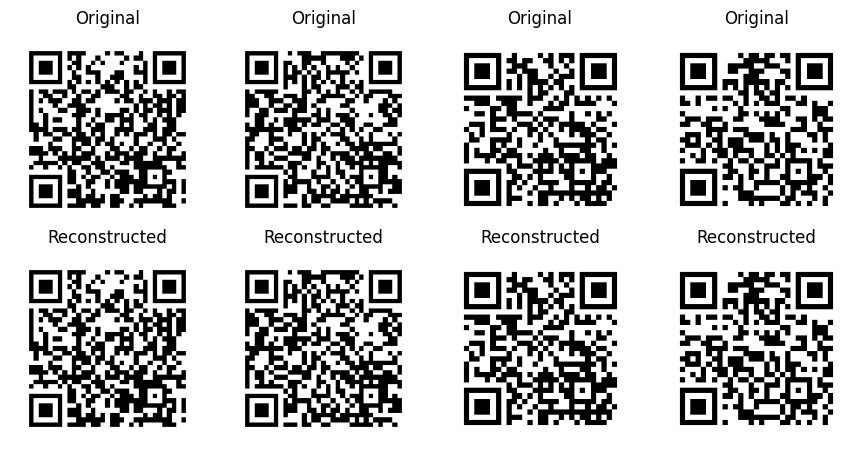}
    \caption{Comparison of original QR codes and their autoencoder reconstructions.}
    \label{fig:ae-reconstruction}
\end{figure}

A final \(3 \times 3\) convolution with one filter and sigmoid activation produces the reconstructed grayscale image of size \(224 \times 224 \times 1\). As shown in Figure \ref{fig:ae-reconstruction}, the reconstructed images preserve the main QR code structure, including the module arrangement and finder patterns. As an initial qualitative check, the original and reconstructed QR codes were scanned using both iPhone and Samsung smartphone cameras. For the evaluated examples, the reconstructed QR codes were successfully recognized and opened the same embedded URLs as their corresponding original images. To provide a quantitative evaluation over the complete test set, the original and reconstructed QR code images were also decoded programmatically using ZXing-C++ and their payloads were compared directly. This evaluation was performed on all test images. The results, reported in Table \ref{tab:qr_decode_results}, show that the reconstructed QR codes preserved the exact payload in 99.73\% of all test images and in 99.76\% of the originally decodable images. Among the 29,989 originally decodable QR codes, only 71 reconstructions could not be decoded. However, no case was observed in which both the original and reconstructed images were successfully decoded but returned different payloads. Together, the qualitative and quantitative results show that the autoencoder preserved both the main visual structure and the functional content of the QR codes. The reconstruction task therefore allowed the encoder to learn important QR specific features without using class labels. After pretraining, the decoder was removed and was not used during classification.

\begin{table}[!htbp]
\centering
\caption{QR code decoding and payload preservation results on the test set.}
\label{tab:qr_decode_results}

\begin{tabular}{lccc}
\hline\noalign{\smallskip}
\textbf{Evaluation Metric} & \textbf{Overall} & \textbf{Benign} & \textbf{Malicious} \\
\noalign{\smallskip}\hline\noalign{\smallskip}
Test images & 30,000 & 15,000 & 15,000 \\
Original QR codes successfully decoded & 29,989 & 15,000 & 14,989 \\
Reconstructed QR codes successfully decoded & 29,918 & 14,998 & 14,920 \\
Exact payload matches & 29,918 & 14,998 & 14,920 \\
Original QR decoding rate (\%) & 99.96 & 100.00 & 99.93 \\
Reconstructed QR decoding rate (\%) & 99.73 & 99.99 & 99.47 \\
Exact payload match rate (\%) & 99.73 & 99.99 & 99.47 \\
Payload retention among decodable originals (\%) & 99.76 & 99.99 & 99.54 \\
Original decoded, reconstruction failed & 71 & 2 & 69 \\
Original failed, reconstruction decoded & 0 & 0 & 0 \\
Both decoded, payload mismatch & 0 & 0 & 0 \\
Neither QR code decoded & 11 & 0 & 11 \\
\noalign{\smallskip}\hline
\end{tabular}
\end{table}

For classification, global average pooling converts the
\(14 \times 14 \times 384\) encoder output into a 384 dimensional
feature vector. Global average pooling reduces the number of
classification parameters while summarizing the activation of each
feature channel. The resulting vector is passed through a dense layer
with 128 units, followed by batch normalization, ReLU activation and
dropout with a rate of \(0.4\). A final dense layer with one sigmoid
produces the probability that the input QR code is benign/malicious.

The QRNet development process begins with autoencoder pretraining and is
followed by two supervised training stages without changing the model
architecture. In Stage 1, the pretrained encoder is frozen and only the
classification head is trained using the benign and malicious labels.
In Stage 2, the encoder convolutional layers and the classification head
are fine-tuned together, allowing the convolutional filters to adapt to
the quishing detection task. The final fine-tuned QRNet model is then
used for the main classification evaluation, robustness testing,
explainability analysis, QR specific transfer tasks, and external dataset
evaluation.

\subsubsection{Hyperparameter Settings}

Hyperparameter tuning was performed using the validation set, while the
test set was kept untouched for the final evaluation. An extensive
randomized search was carried out for the ML and
DL models, as well as QRNet, using wide predefined
hyperparameter ranges. For each model, many hyperparameter combinations
were tested to find the configuration with the best validation
performance based on ROC-AUC. Table \ref{tab:baseline-hyperparameters} presents the best performing hyperparameter combination selected for
each of the 12 baseline models and QRNet to support reproducibility.

\begin{table}[!htbp]
\centering
\caption{Hyperparameters Used for the Evaluated Models.}
\label{tab:baseline-hyperparameters}

\begingroup
\normalsize
\setlength{\tabcolsep}{5pt}
\renewcommand{\arraystretch}{1.20}

\begin{tabularx}{\textwidth}{@{}
>{\raggedright\arraybackslash}p{0.22\textwidth}
>{\raggedright\arraybackslash}X@{}}

\hline\noalign{\smallskip}
\textbf{Model} & \textbf{Selected hyperparameters} \\
\noalign{\smallskip}\hline\noalign{\smallskip}

\multicolumn{2}{@{}l}{\textit{Machine learning baselines}} \\
\noalign{\smallskip}

Logistic Regression & Solver: sag; penalty: $\ell_2$; class weight: none; $C=10$ \\

SVM & Class weight: none; $C=3$ \\

Naive Bayes & Fit prior: false; binarize: 0.1; $\alpha=0.5$ \\

Random Forest & Number of estimators: 300; maximum features: \texttt{sqrt}; maximum depth: 30 \\

LightGBM & Number of leaves: 127; number of estimators: 300; learning rate: 0.1 \\

XGBoost & Number of estimators: 300; maximum depth: 10; learning rate: 0.1 \\

CatBoost & Learning rate: 0.1; iterations: 800; depth: 8 \\

AdaBoost & Number of estimators: 100; learning rate: 0.1; base-estimator maximum depth: 2 \\[3pt]

\multicolumn{2}{@{}l}{\textit{Deep learning baselines}} \\
\noalign{\smallskip}

EfficientNetB0 & Dense units: 256; head dropout: 0.2; initial learning rate: 0.0003;
batch size: 32; optimizer: Adam; best learning rate: $3\times10^{-5}$ \\

ResNet50 & Dense units: 256; head dropout: 0.2; initial learning rate: 0.0001;
batch size: 32; optimizer: Adam; best learning rate: $2\times10^{-5}$ \\

DenseNet121 & Dense units: 128; head dropout: 0.2; initial learning rate: 0.0003;
batch size: 32; optimizer: Adam; best learning rate: $2\times10^{-5}$ \\

YOLOv11 & Model: yolo11m-cls.pt; task: classification; optimizer: AdamW;
learning rate: 0.0001; weight decay: 0.0005; batch size: 32 \\[3pt]

\multicolumn{2}{@{}l}{\textit{Proposed model}} \\
\noalign{\smallskip}

QRNet & Base filters: 48; dense units: 128; latent space: 384; head dropout: 0.4;
optimizer: Adam; learning rate: 0.001; batch size: 32 \\

\noalign{\smallskip}\hline

\end{tabularx}
\endgroup
\end{table}

\subsection{Evaluation Metrics and Statistical Tests}

\subsubsection{Evaluation Metrics}

All models were evaluated on the same test set using accuracy, precision, recall, F1-score and ROC-AUC. In this study, malicious QR images were treated as the positive class, while benign QR images were treated as the negative class.

\textbf{Accuracy} measures the proportion of all correctly classified samples. It provides an overall measure of how often the model produces
the correct prediction.

\begin{equation}
\text{Accuracy} =
\frac{TP+TN}{TP+TN+FP+FN}.
\end{equation}

\textbf{Precision} measures the proportion of QR images predicted as malicious that were actually malicious. A high precision value means that the model produces fewer false alarms for benign QR images.

\begin{equation}
\text{Precision} =
\frac{TP}{TP+FP}.
\end{equation}

\textbf{Recall} measures the proportion of malicious QR images that were correctly detected. A high recall value means that the model misses fewer malicious QR images.

\begin{equation}
\text{Recall} =
\frac{TP}{TP+FN}.
\end{equation}

\textbf{F1-score} is the harmonic mean of precision and recall. It provides a balanced measure of performance when both false positives and false negatives are important.

\begin{equation}
\text{F1-score} =
2 \times
\frac{\text{Precision}\times\text{Recall}}
{\text{Precision}+\text{Recall}}.
\end{equation}

\textbf{ROC-AUC} measures the ability of a model to distinguish between benign and malicious QR images across different classification thresholds. The receiver operating characteristic curve plots the true positive rate against the false positive rate. An ROC-AUC value
close to 1 indicates strong class separation, while a value close to 0.5 indicates performance similar to random classification.

\begin{equation}
\text{TPR} =
\frac{TP}{TP+FN},
\end{equation}

\begin{equation}
\text{FPR} =
\frac{FP}{FP+TN}.
\end{equation}

The area under the ROC curve is calculated as:

\begin{equation}
\text{ROC-AUC} =
\int_{0}^{1}
\text{TPR}(\text{FPR})\, d(\text{FPR}).
\end{equation}

\subsubsection{Statistical Tests}

Although evaluation metrics show which model achieved higher scores,
small numerical differences may not be meaningful. Therefore, the
DeLong and McNemar tests were used to determine whether the
differences between QRNet and the strongest evaluated models were
statistically significant rather than only numerical. Each comparison
was performed using the same paired test samples for both models. A
significance level of $\alpha = 0.05$ was used for both tests.

\textbf{DeLong Test} was used to compare the ROC-AUC values of QRNet and
each baseline model. This test is suitable for paired model comparisons
because both models were evaluated on the same test samples. Their ROC
curves are therefore correlated. The test uses the continuous prediction
scores produced by the models rather than the binary predictions obtained
from a fixed threshold.

The null hypothesis states that the two models have equal ROC-AUC
values, while the alternative hypothesis states that their ROC-AUC
values are different:

\begin{equation}
H_0:
\mathrm{AUC}_{\mathrm{QRNet}}
=
\mathrm{AUC}_{\mathrm{baseline}},
\end{equation}

\begin{equation}
H_1:
\mathrm{AUC}_{\mathrm{QRNet}}
\neq
\mathrm{AUC}_{\mathrm{baseline}}.
\end{equation}

The difference between the two estimated ROC-AUC values is defined as:

\begin{equation}
d =
\widehat{\mathrm{AUC}}_{\mathrm{QRNet}}
-
\widehat{\mathrm{AUC}}_{\mathrm{baseline}}.
\end{equation}

The DeLong test estimates the variance of this difference while taking
the correlation between the two ROC curves into account. The test
statistic is calculated as:

\begin{equation}
z =
\frac{d}
{\sqrt{\mathrm{Var}(d)}}.
\end{equation}

The two sided $p$-value is then calculated as:

\begin{equation}
p =
2\left(1-\Phi\left(|z|\right)\right),
\end{equation}

where $\Phi$ is the cumulative distribution function of the standard
normal distribution. A $p$-value below 0.05 leads to the rejection of
the null hypothesis and indicates that the difference between the two
ROC-AUC values is statistically significant.

\textbf{McNemar Test} was used to compare the binary classification errors
made by QRNet and each baseline model on the same test samples. Unlike
the DeLong test, which uses prediction scores to compare ROC-AUC
values, the McNemar test uses the final class predictions obtained at
a threshold of 0.5. It considers only discordant samples, for which one
model produced a correct prediction and the other produced an incorrect
prediction.

Let $b$ represent the number of samples incorrectly classified by
QRNet but correctly classified by the baseline model. Let $c$ represent
the number of samples correctly classified by QRNet but incorrectly
classified by the baseline model. Samples classified correctly by both
models or incorrectly by both models do not contribute directly to the
test.

Let $\pi_b$ denote the probability of a discordant sample belonging to
the first case and $\pi_c$ denote the probability of it belonging to the
second case. Under the null hypothesis, the two disagreement directions
are equally likely:

\begin{equation}
H_0: \pi_b = \pi_c = 0.5.
\end{equation}

The alternative hypothesis states that the two probabilities are
different:

\begin{equation}
H_1: \pi_b \neq \pi_c.
\end{equation}

Let $B$ be the random variable representing the number of disagreement
samples of type $b$. Conditional on the total number of discordant
samples, $n=b+c$, the null distribution is:

\begin{equation}
B \sim \mathrm{Binomial}(b+c,0.5).
\end{equation}

The observed test statistic is the smaller of the two disagreement
counts:

\begin{equation}
k=\min(b,c).
\end{equation}

The two sided $p$-value is calculated as:

\begin{equation}
p =
\min\left\{
1,\,
2P\left(B\leq k\right)
\right\}.
\end{equation}

A $p$-value below 0.05 leads to the rejection of the null hypothesis and
indicates that the two models have significantly different
classification error probabilities. When $c>b$, QRNet correctly
classifies more discordant samples than the baseline model, indicating
that QRNet makes fewer classification errors on the paired test set.

\subsection{Robustness and Explainability}

\subsubsection{Robustness Evaluation}

Robustness was evaluated using QRNet model, which was
the best performing model in this study. The evaluation used the full
test set of 150,861 QR code images. The model weights were kept
fixed and no distorted images were used during training or
fine-tuning. All images were converted to grayscale, resized to
\(224 \times 224\) and normalized to the range \([0,1]\). The same
classification threshold of 0.5 was used for both clean and distorted
images. The model was not retrained and the threshold was not changed
for any condition.

In addition to the clean baseline, QRNet was tested under 27 distorted
conditions.  Blur was applied using radius values of \(0.3\), \(0.5\),
\(0.8\), \(1.0\) and \(2.0\). Gaussian noise was added using standard
deviations of \(0.05\), \(0.10\), \(0.15\) and \(0.20\). JPEG
compression was tested at quality levels of \(90\), \(70\), \(50\) and
\(30\). Brightness and contrast were changed using factors of \(0.5\) and \(0.7\). Resolution changes were tested by first resizing the images from \(224 \times 224\) to lower resolutions of \(192 \times 192\), \(168 \times 168\) and \(112 \times 112\) or to higher resolutions of \(336 \times 336\), \(448 \times 448\) and \(672 \times 672\). The images were then resized back to \(224 \times 224\).
Random white square occlusions covering about \(5\%\), \(10\%\), \(15\%\), and \(20\%\) of the image area were also applied. A fixed random seed was used for each noise and occlusion condition to make the experiment reproducible.

Each distortion was applied separately to the full test set.
Performance was measured using accuracy, precision,
recall, F1-score, macro-F1 and ROC-AUC. The effect of each distortion
was measured by comparing its macro-F1 and ROC-AUC values with the
values obtained on the clean test set:

\begin{equation}
\Delta F1_{\mathrm{macro},c}
=
F1_{\mathrm{macro,clean}}
-
F1_{\mathrm{macro},c},
\end{equation}

\begin{equation}
\Delta \mathrm{AUC}_{c}
=
\mathrm{AUC}_{\mathrm{clean}}
-
\mathrm{AUC}_{c},
\end{equation}

where \(c\) represents a distorted condition. A positive value means
that performance decreased compared with the clean test set.

\subsubsection{Explainability Methods}

Explainability methods were applied to QRNet and XGBoost because they
were the two highest performing models in this study. QRNet learns
directly from QR code images, while XGBoost uses extracted HOG features.
The aim was to examine the features and image regions used by each model
when making predictions. Local methods were used to explain individual
predictions, while global methods were used to study feature importance
across a larger set of test samples.

For QRNet, five local explainability methods were applied to
test images containing high confidence and borderline predictions from
both classes. Grad-CAM used the final encoder output of size
\(14 \times 14 \times 384\) to highlight the main regions that influenced
each prediction. Occlusion sensitivity divided each image into a
\(14 \times 14\) grid and covered one region at a time with white pixels.
The change in the predicted probability was then used to measure the
importance of each region. Integrated Gradients used a white image as
the baseline and 64 integration steps to measure the contribution of
each pixel. LIME divided each image into 80 regions of similar pixels and generated 1,000 modified samples by hiding different regions. Gradient
SHAP used a balanced background set to produce attribution maps for the
selected images. These methods provided different views of the regions
that influenced QRNet predictions.

Global explainability for QRNet was performed using the 384 latent
features produced by the global average pooling layer. Gradient SHAP was
applied to a balanced set of 10,000 test images, containing 5,000 benign
and 5,000 malicious samples. The mean absolute SHAP value was used to
measure the overall importance of each latent feature.

For XGBoost, TreeSHAP was applied using \texttt{TreeExplainer} to explain
the selected HOG features. SHAP values were calculated using a balanced
set of 10,000 test samples, with 5,000 samples from each class. The mean
absolute SHAP value was used to rank the overall importance of the HOG
features. The selected features were also mapped back to their related
image cells to produce spatial importance maps over the QR code images.

Finally, the heatmaps of the image-based QRNet model and the
HOG feature-based XGBoost model were compared to see whether both models
focused on the same regions of the QR codes when making predictions.

\subsection{Generalization and Transfer Evaluation}

\subsubsection{QR Specific Transfer Tasks}

Two additional multiclass tasks were used to examine whether the
pretrained QRNet model learned structural information about QR codes
beyond benign and malicious classification. The first task classified
the QR version, while the second classified the QR mask pattern. Both
tasks used structural labels obtained from the same image collection as
the main study. Therefore, they were treated as QR specific transfer
tasks rather than external generalization tests.

\textbf{QR Version and Mask Pattern Labeling.}
QR version labels were obtained from the number of modules along each
side of the QR code. The estimated grid size was matched to the nearest
valid QR version. Images that could not be labeled reliably were
removed. Versions with very few samples were also excluded. The final
task included 13 QR versions: \(1, 2, 3, 4, 5, 6, 7, 8, 9, 11, 12, 21\) and \(22\). Mask pattern labels were obtained from the 15-bit format information field of each QR code. The extracted information was compared with the 32 valid format codewords, which represent four error correction levels and eight mask patterns. The closest valid codeword was accepted when its Hamming distance was no greater than three. Images without a reliable mask pattern label were excluded. The task included all 8 QR mask patterns, numbered from \(0\) to \(7\).

\textbf{Stage 1: Frozen Encoder Transfer.}
In the first stage, the pretrained QRNet encoder was kept frozen and a
new multiclass classification head was trained for each task. The
version task used 13 output classes, while the mask pattern task used
8 output classes. Both tasks used around 200,000 labeled QR code images. The same general training procedure and main hyperparameter settings used for the original QRNet model were followed, with a different output layer for each task.

\textbf{Stage 2: Encoder Fine-Tuning.}
In the second stage, the convolutional encoder layers and the
classification head were fine-tuned together. Version classification
used 956,386 labeled images, while mask pattern classification used
956,330 labeled images. The same general hyperparameters and
fine-tuning procedure used for QRNet were followed.

Both tasks were evaluated using the same metrics as the main QRNet
model: accuracy, precision, recall, F1-score, macro-F1 and ROC-AUC.

\subsubsection{External Dataset Evaluation}

The external evaluation used the QR Code Quishing 200k dataset
\cite{ahsan2026qrcodequishing200k}, which contains 200,000 QR code
images, including 100,000 benign and 100,000 malicious images. The
dataset was divided through stratified sampling into a training set of
140,000 images, a validation set of 30,000 images and a test set of
30,000 images. Each split contained an equal number of benign and
malicious images. All images were converted to grayscale, resized to
\(224 \times 224\) and normalized to the range \([0,1]\). The test set
was kept unseen during training and validation.

\textbf{Zero-shot evaluation} was performed on the QR Code Quishing 200k dataset \cite{ahsan2026qrcodequishing200k}. The QRNet model was
applied directly to the external images without any retraining, changing the classification threshold, tuning the hyperparameters or selecting model weights based on the external validation set. The model was evaluated directly on the complete 200,000 image dataset and on the fixed 30,000 image test set. Evaluation on the test set also provided a direct baseline for comparison with the fine-tuned model.

\textbf{Fine-tuning} was performed on the QR Code Quishing 200k dataset
\cite{ahsan2026qrcodequishing200k}. The original QRNet model was
reloaded and fine-tuned using the training set. The validation set was
used to identify the training epoch with the highest validation
ROC--AUC and the model weights from that epoch were restored before
evaluation on the unseen test set. The QRNet fine-tuning procedure and
hyperparameter settings reported in Table
\ref{tab:baseline-hyperparameters} were used for this experiment.

A fixed classification threshold of \(0.5\) was used throughout the
external evaluation. Performance was measured using accuracy,
precision, recall, F1-score, macro-F1 and ROC--AUC. These experiments evaluated the zero-shot across dataset transfer of QRNet and its ability to adapt to external dataset through supervised fine-tuning.

\begin{figure}[!htbp]
    \centering
    \includegraphics[width=\textwidth]{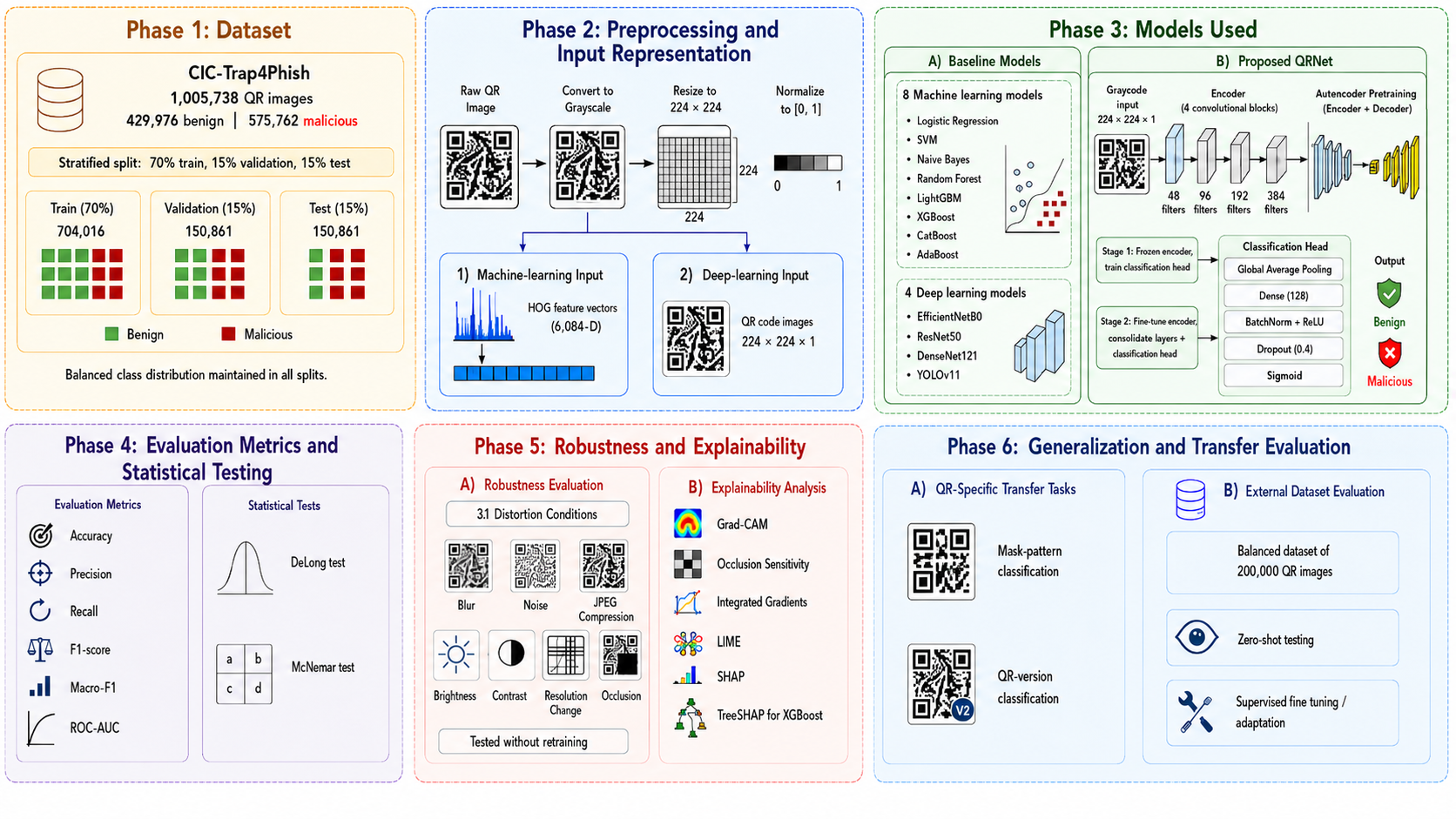}
    \caption{Overview of the methodology followed in this study.}
    \label{fig:methodology-overview}
\end{figure}

Figure \ref{fig:methodology-overview} summarizes the complete methodology of this study, from dataset organization and preprocessing to model evaluation, robustness, explainability analysis, generalization
and transfer evaluation.

\section{Results and Discussion}
\label{sec:results}

This section presents the classification results of the evaluated
models, followed by statistical validation of QRNet. Then, it examines
the robustness and explainability of the proposed model and reports the
results of the QR specific transfer tasks and external dataset
evaluation.

\subsection{Classification Performance and Statistical Validation}

\subsubsection{Classification Performance}

The classification experiment compared 13 models under the same
setting. These included 8 ML models trained using HOG
features, 4 ImageNet-pretrained DL models, and the proposed
QRNet model. All models were evaluated on the same unseen
test set from the balanced 200,000 image subset using accuracy,
precision, recall, F1-score, macro-F1 and ROC--AUC.

\begin{figure}[!htbp]
    \centering
    \includegraphics[width=0.9\textwidth]{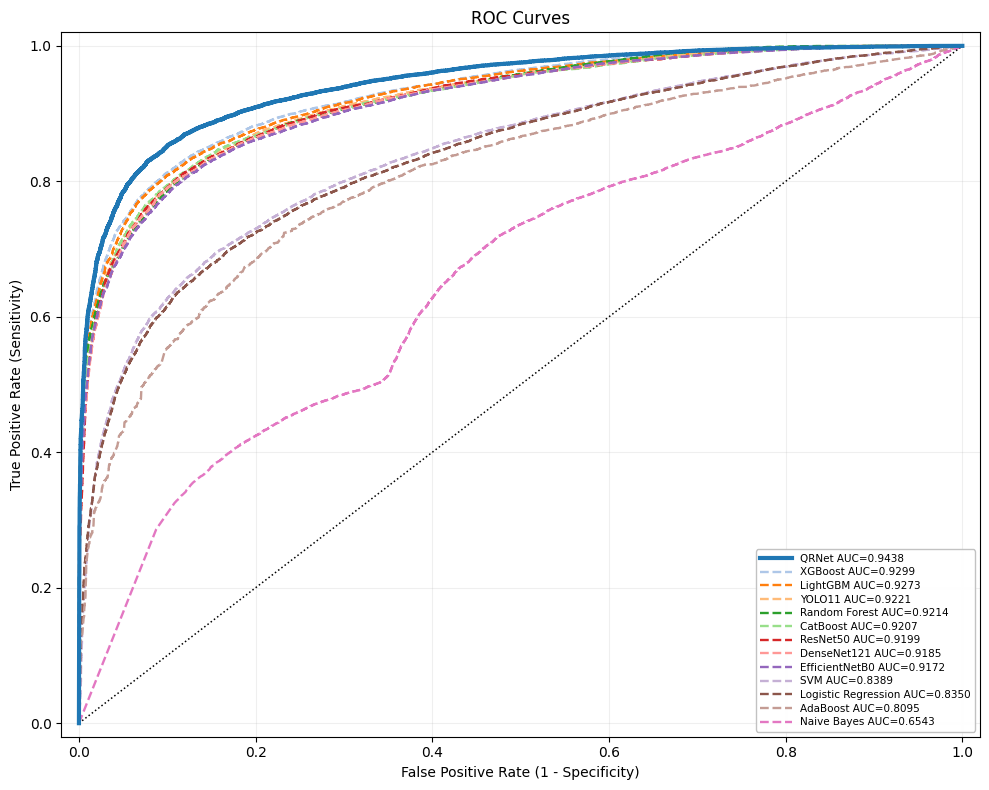}
    \caption{ROC Curves of ML and DL Models.}
    \label{fig:roc-curves}
\end{figure}

Table \ref{tab:ml-malicious-test} presents the numerical results, while
Figure \ref{fig:roc-curves} compares the ROC curves of the evaluated
models. Among the ML models, XGBoost achieved the strongest performance, with an accuracy of \(0.8556\), an F1-score of \(0.8475\), a macro-F1 score of \(0.8552\) and a ROC--AUC of \(0.9298\). LightGBM and CatBoost also performed well, showing that tree-based ensemble methods can capture useful nonlinear relations among the HOG features. Naive Bayes produced the lowest results, as its independence assumption is not well suited to HOG features whose neighboring gradient values may be related. Among the DL models, YOLOv11 achieved the best results, with an accuracy of \(0.8464\), an F1-score of \(0.8414\), a macro-F1 score of \(0.8463\) and a ROC--AUC of \(0.9221\). DenseNet121 achieved higher precision, while ResNet50 achieved higher recall, but none of the four DL models exceeded XGBoost in overall performance.

QRNet achieved the best results among the 13 models, achieving an
accuracy of \(0.8754\), a recall of \(0.8422\), an F1-score of
\(0.8711\), a macro-F1 score of \(0.8753\) and a ROC--AUC of
\(0.9437\). Compared with XGBoost, which was the second best overall
model, QRNet increased accuracy from \(0.8556\) to
\(0.8754\), F1-score from \(0.8475\) to \(0.8711\), macro-F1 from
\(0.8552\) to \(0.8753\) and ROC--AUC from \(0.9298\) to \(0.9437\).

Training from scratch on QR code images produced better results than
using HOG-based ML models or ImageNet pretrained DL models. ImageNet is a large object centered dataset containing
general visual categories that differ from regular module patterns
and structural details of QR codes \cite{5206848,Russakovsky2015}. In contrast, QRNet learned its features directly from the target image domain. These findings show that QR specific training is more suitable and effective for quishing attack detection than HOG-based
feature engineering or transfer learning from ImageNet pretrained models.

\begin{table}[!htbp]
\centering
\caption{Performance of the evaluated models on the test set.}
\label{tab:ml-malicious-test}

\begingroup
\normalsize
\setlength{\tabcolsep}{4pt}
\renewcommand{\arraystretch}{1.15}
\sisetup{detect-weight=true}

\begin{tabular*}{\textwidth}{@{\extracolsep{\fill}}l *{6}{S[table-format=1.4]}@{}}

\hline\noalign{\smallskip}
\textbf{Model} & {\textbf{Accuracy}} & {\textbf{Precision}} &
{\textbf{Recall}} & {\textbf{F1-score}} & {\textbf{Macro-F1}} &
{\textbf{ROC-AUC}} \\
\noalign{\smallskip}\hline\noalign{\smallskip}

\multicolumn{7}{@{}l}{\textit{Machine learning baselines}} \\[2pt]

Logistic Regression & 0.7613 & 0.7804 & 0.7273 & 0.7529 & 0.7610 & 0.8350 \\
SVM                 & 0.7657 & 0.7884 & 0.7262 & 0.7560 & 0.7653 & 0.8388 \\
Naive Bayes         & 0.5876 & 0.6028 & 0.5136 & 0.5547 & 0.5853 & 0.6598 \\
Random Forest       & 0.8385 & 0.9055 & 0.7558 & 0.8239 & 0.8373 & 0.9214 \\
LightGBM            & 0.8542 & 0.8935 & 0.8043 & 0.8465 & 0.8538 & 0.9272 \\
XGBoost             & 0.8556 & 0.8973 & 0.8030 & 0.8475 & 0.8552 & 0.9298 \\
CatBoost            & 0.8469 & 0.8833 & 0.7990 & 0.8393 & 0.8465 & 0.9206 \\
AdaBoost            & 0.7420 & 0.7762 & 0.6802 & 0.7250 & 0.7411 & 0.8094 \\

\noalign{\smallskip}
\multicolumn{7}{@{}l}{\textit{Deep learning baselines}} \\[2pt]

EfficientNetB0      & 0.8414 & 0.8616 & 0.8135 & 0.8368 & 0.8413 & 0.9171 \\
ResNet50            & 0.8446 & 0.8586 & 0.8250 & 0.8414 & 0.8445 & 0.9198 \\
DenseNet121         & 0.8453 & 0.8785 & 0.8013 & 0.8381 & 0.8450 & 0.9184 \\
YOLOv11             & 0.8464 & 0.8699 & 0.8147 & 0.8414 & 0.8463 & 0.9221 \\

\noalign{\smallskip}\hline\noalign{\smallskip}

\textbf{QRNet}      & \bfseries 0.8754 & \bfseries 0.9022 & \bfseries 0.8422 &
                       \bfseries 0.8711 & \bfseries 0.8753 & \bfseries 0.9437 \\

\noalign{\smallskip}\hline

\end{tabular*}
\endgroup
\end{table}

\begin{figure}[!htbp]
    \centering
    \includegraphics[width=\textwidth]{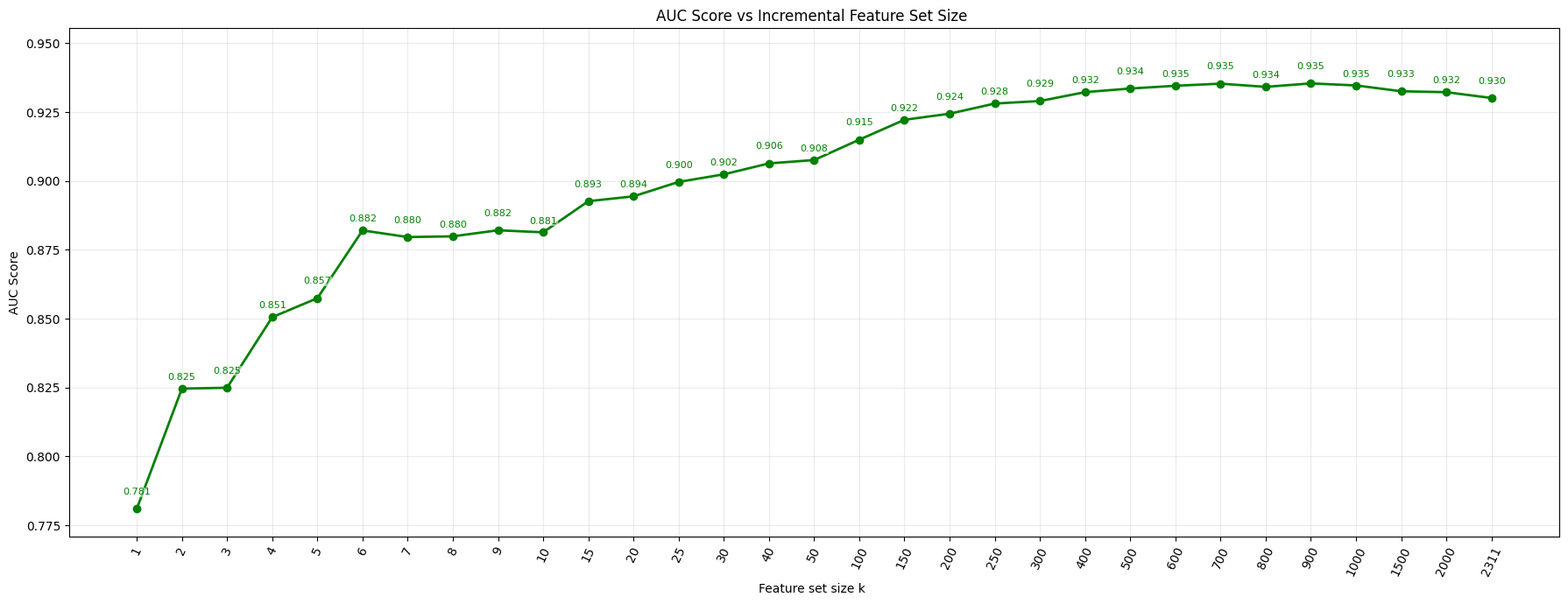}
    \caption{XGBoost ROC--AUC Across Incremental HOG Feature Set Sizes}
    \label{fig:xgboost-hog-feature-selection}
\end{figure}

Before training XGBoost on the complete dataset, the original
6,084 dimensional HOG representation was reduced through gain-based
feature selection. The features were ranked according to their XGBoost
gain importance and the 2,311 features with non-zero gain were retained
for incremental evaluation. A new XGBoost model was trained for each
top-\(k\) subset using the same hyperparameter configuration.

As shown in Figure \ref{fig:xgboost-hog-feature-selection}, ROC--AUC
increased as the highest ranked HOG features were added and began to
plateau after several hundred features. The subset containing the top
700 features achieved an ROC--AUC of \(0.9353\), exceeding the
\(0.9298\) obtained using the original representation of 6,084 HOG features. It was also the smallest evaluated subset to exceed an ROC--AUC of \(0.935\). Therefore, the 700-feature subset was selected for training
XGBoost on the complete dataset, reducing the original HOG
dimensionality by approximately \(88.5\%\) while improving
classification performance.

\begin{table}[!htbp]
\centering
\caption{Performance comparison of XGBoost and QRNet on the test set ($n=150{,}861$)}
\label{tab:million-image-comparison}

\begingroup
\normalsize
\setlength{\tabcolsep}{4pt}
\renewcommand{\arraystretch}{1.15}
\sisetup{detect-weight=true}

\begin{tabular*}{\textwidth}{@{\extracolsep{\fill}}l *{6}{S[table-format=1.4]}@{}}

\hline\noalign{\smallskip}
\textbf{Model} & {\textbf{Accuracy}} & {\textbf{Precision}} &
{\textbf{Recall}} & {\textbf{F1-score}} & {\textbf{Macro-F1}} &
{\textbf{ROC-AUC}} \\
\noalign{\smallskip}\hline\noalign{\smallskip}

XGBoost (700 features) & 0.8822 & 0.9197 & 0.8702 & 0.8943 & 0.8806 & 0.9537 \\[2pt]

\textbf{QRNet} & \bfseries 0.9052 & \bfseries 0.9323 & \bfseries 0.8997 &
                  \bfseries 0.9157 & \bfseries 0.9037 & \bfseries 0.9680 \\

\noalign{\smallskip}\hline

\end{tabular*}
\endgroup
\end{table}

After selecting the top 700 HOG features, XGBoost was retrained on the
full dataset of 1,005,738 QR code images using the same training,
validation and test sets as QRNet. Table \ref{tab:million-image-comparison}
compares both models on the same unseen test set of 150,861 QR code
images. Although the optimized XGBoost model achieved strong results,
QRNet outperformed it across all metrics. QRNet achieved an
accuracy of \(90.52\%\), a precision of \(93.23\%\), a recall of
\(89.97\%\), an F1-score of \(91.57\%\), a macro-F1 score of
\(90.37\%\) and a ROC--AUC of \(96.80\%\). Therefore, even after
reducing the HOG representation to its strongest 700 feature subset,
QRNet remained the best performing model. The statistical significance
of this improvement over XGBoost is examined in the following
subsection.

\subsubsection{Statistical Validation}

Paired statistical tests were used to determine whether the observed
performance improvements of QRNet over XGBoost and YOLO11m-cls were
statistically significant. Statistical significance was assessed at
\(p<0.05\). Table \ref{tab:statistical-validation} summarizes the
DeLong and McNemar test results.

In comparison with XGBoost using the selected 700 HOG features, QRNet
achieved an ROC--AUC of \(0.968243\), compared with \(0.954131\) for
XGBoost. This represents an AUC improvement of \(0.014112\). The
DeLong test confirmed that this improvement was statistically
significant (\(z=14.40\), \(p=5.36\times10^{-47}\)). Moreover, the
paired bootstrap confidence interval for the AUC difference was narrow
and remained entirely above zero, indicating a precise and consistent
improvement in favor of QRNet. At the fixed threshold of \(0.5\), QRNet
correctly classified 1,314 images that XGBoost misclassified, whereas
XGBoost correctly classified 792 images that QRNet misclassified. This
corresponded to a \(2.32\) percentage point reduction in classification
error. The exact McNemar test (\(p=3.66\times10^{-30}\)) and the
confidence interval for the paired error rate reduction confirmed that
QRNet produced significantly fewer errors.

In comparison with YOLO11m-cls, QRNet achieved a ROC--AUC of
\(0.943788\), compared with \(0.922146\) for YOLO11m-cls, representing
an improvement of \(0.021642\). The DeLong test showed that this
difference was statistically significant
(\(z=17.90\), \(p=1.18\times10^{-71}\)). The paired bootstrap
confidence interval was narrow and remained entirely above zero,
demonstrating a consistent improvement in favor of QRNet. At the fixed
threshold, QRNet correctly classified 2,242 images that YOLO11m-cls
misclassified, whereas YOLO11m-cls correctly classified 1,372 images
that QRNet misclassified. This corresponded to a
\(2.90\) percentage point reduction in classification error. The exact
McNemar test (\(p=8.29\times10^{-48}\)) and the confidence interval for
the paired error rate reduction further confirmed that QRNet produced
significantly fewer errors.

Based on these results, QRNet achieved numerical improvements over the
strongest ML and DL models, and these
improvements were statistically significant.

\begin{table}[!htbp]
\centering
\caption{Statistical comparison of QRNet with XGBoost and YOLO11m-cls.}
\label{tab:statistical-validation}

\begingroup
\small
\setlength{\tabcolsep}{4pt}
\renewcommand{\arraystretch}{1.10}

\resizebox{\textwidth}{!}{%
\begin{tabular}{@{}llcll@{}}

\hline\noalign{\smallskip}
\textbf{Comparison} &
\textbf{Test} &
\textbf{Result} &
\textbf{\(\boldsymbol{p}\)-value} &
\textbf{Conclusion} \\
\noalign{\smallskip}\hline\noalign{\smallskip}

QRNet vs. XGBoost-700 &
DeLong &
\(\Delta\mathrm{AUC}=0.014112,\; z=14.40\) &
\(5.36\times10^{-47}\) &
QRNet achieved a higher AUC \\

QRNet vs. XGBoost-700 &
McNemar &
\(792\) vs. \(1,314\) discordant cases &
\(3.66\times10^{-30}\) &
QRNet produced fewer errors \\

\noalign{\smallskip}\hline\noalign{\smallskip}

QRNet vs. YOLO &
DeLong &
\(\Delta\mathrm{AUC}=0.021642,\; z=17.90\) &
\(1.18\times10^{-71}\) &
QRNet achieved a higher AUC \\

QRNet vs. YOLO &
McNemar &
\(1,372\) vs. \(2,242\) discordant cases &
\(8.29\times10^{-48}\) &
QRNet produced fewer errors \\

\noalign{\smallskip}\hline

\end{tabular}%
}

\par\vspace{0.3em}
{\footnotesize\raggedright
For the McNemar results, discordant counts are reported as QRNet incorrect and comparator correct, followed by QRNet correct and comparator incorrect.\par}

\endgroup
\end{table}

\subsection{Robustness and Explainability Analysis}

\subsubsection{Robustness Under Image Distortions}

\begin{figure}[!htbp]
    \centering
    \includegraphics[
        width=0.75\textwidth,
        height=0.50\textheight,
        keepaspectratio
    ]{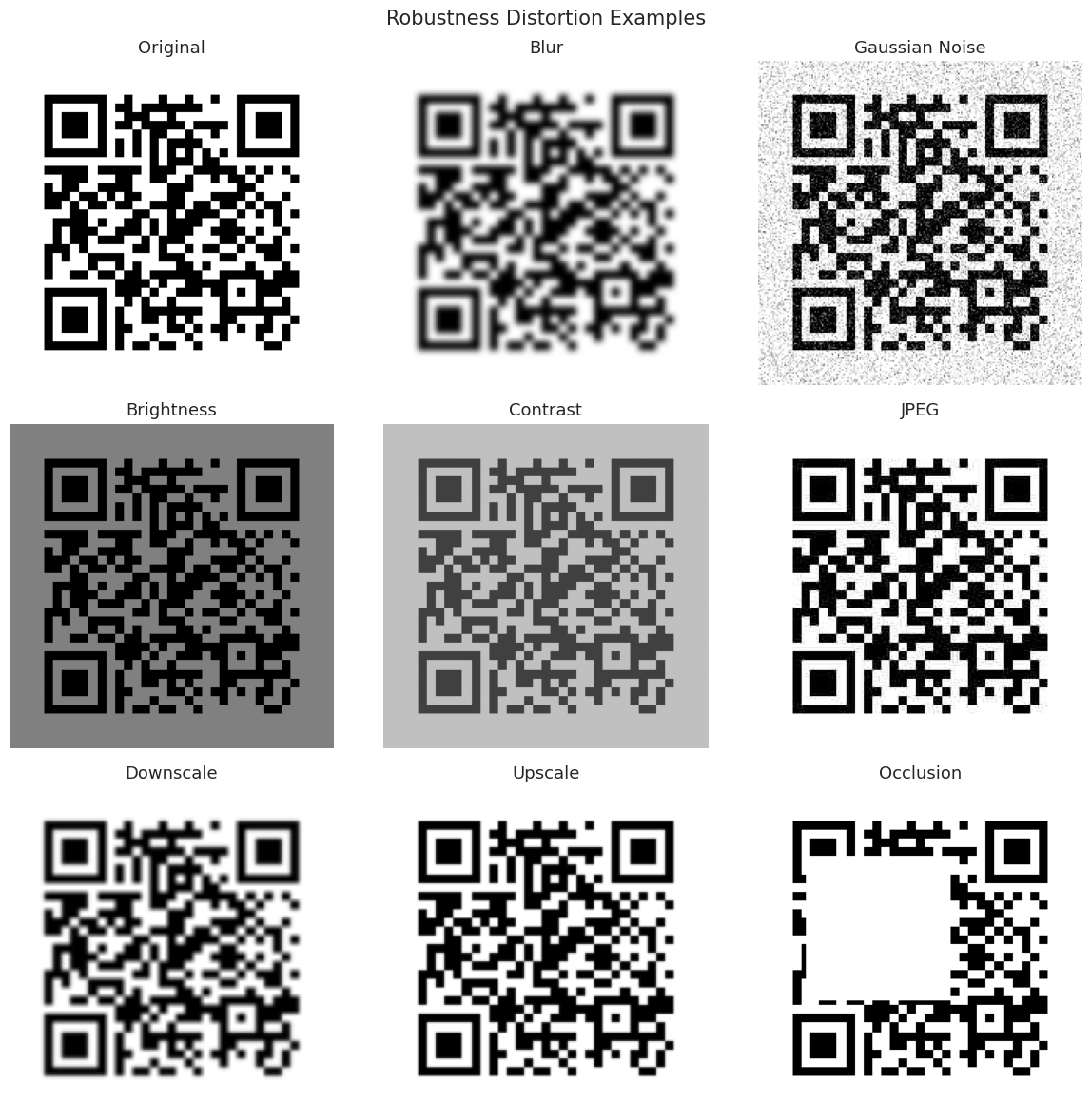}
    \caption{Examples of QR code distortions in QRNet robustness evaluation.}
    \label{fig:robustness-distortions}
\end{figure}

To assess how QRNet performs when QR code quality changes, the final model
was evaluated without retraining under 27 distortion conditions on the whole test set. The results are reported in Table \ref{tab:robustness-distortions}, while Figure \ref{fig:robustness-distortions} shows some examples of the applied distortions. QRNet remained reliable when the main QR structure was preserved, but its performance decreased when module boundaries became unclear or large parts of the code were removed.

The strongest results were obtained under Gaussian noise, JPEG
compression, brightness and contrast changes, and upscaling. Even
with Gaussian noise at $\sigma=0.20$ and JPEG quality reduced to 30,
macro-F1 and ROC--AUC remained very close to the clean baseline.
Upscaling also caused only small changes. Brightness and contrast factors
of 0.7 were well tolerated, whereas reducing either factor to 0.5 lowered
precision and increased malicious recall. This shows that strong
darkening and contrast loss caused more benign QR codes to be classified
as malicious. 

The largest performance losses occurred under severe blur and
aggressive downscaling. Blur radii of 0.3 and 0.5 caused almost no
degradation, but the decline became clear at radius 0.8 and severe at
radius 1.0. At radius 2.0, the model classified all samples as malicious,
showing that QRNet depends strongly on sharp module edges and fine local
details. Downscaling produced a similar pattern, with performance decreasing at $192\times192$ and $168\times168$, while at $112\times112$ almost all samples were predicted as malicious. In contrast, upscaling remained well tolerated. Occlusion caused a more gradual decline, from a small effect at 5\% occlusion to a much clearer loss at 20\%, indicating that QRNet can handle limited missing information but becomes less reliable
as larger regions of the QR code are removed.

\begin{table}[!htbp]
\centering
\caption{QRNet robustness on clean QR codes and under 27 distortion conditions.}
\label{tab:robustness-distortions}

\begingroup
\small
\setlength{\tabcolsep}{3pt}
\renewcommand{\arraystretch}{1.15}

\resizebox{\textwidth}{!}{%
\begin{tabular}{@{}r l r r r r r r r r r@{}}

\hline\noalign{\smallskip}
& \textbf{Condition} & \textbf{Accuracy} & \textbf{Precision} &
\textbf{Recall} & \textbf{F1-score} & \textbf{MF1} &
\textbf{ROC-AUC} & \textbf{$\Delta$MF1} & \textbf{$\Delta$AUC} &
\textbf{Seconds} \\
\noalign{\smallskip}\hline\noalign{\smallskip}

0  & clean & 0.905251 & 0.932334 & 0.899796 & 0.915776 & 0.903747 & 0.968065 & 0.000000 & 0.000000 & 667.4 \\
1  & blur\_radius\_0.3   & 0.906066 & 0.932350 & 0.901313 & 0.916569 & 0.904553 & 0.968423 & \cellcolor{green!20}$-0.000806$ & \cellcolor{green!20}$-0.000358$ & 848.0 \\
2  & blur\_radius\_0.5   & 0.906225 & 0.926527 & 0.908214 & 0.917279 & 0.904520 & 0.968273 & \cellcolor{green!20}$-0.000773$ & \cellcolor{green!20}$-0.000208$ & 830.2 \\
3  & blur\_radius\_0.8   & 0.860799 & 0.824994 & 0.960620 & 0.887656 & 0.852361 & 0.961883 & \cellcolor{red!20}0.051386 & \cellcolor{red!20}0.006182 & 827.9 \\
4  & blur\_radius\_1.0   & 0.653569 & 0.623477 & 0.996874 & 0.767152 & 0.545396 & 0.949441 & \cellcolor{red!20}0.358351 & \cellcolor{red!20}0.018624 & 826.1 \\
5  & blur\_radius\_2.0   & 0.572474 & 0.572474 & 1.000000 & 0.728119 & 0.364059 & 0.862541 & \cellcolor{red!20}0.539688 & \cellcolor{red!20}0.105523 & 830.8 \\
6  & gaussian\_noise\_0.05 & 0.905575 & 0.932487 & 0.900236 & 0.916078 & 0.904073 & 0.968245 & \cellcolor{green!20}$-0.000326$ & \cellcolor{green!20}$-0.000180$ & 777.7 \\
7  & gaussian\_noise\_0.10 & 0.906026 & 0.931942 & 0.901695 & 0.916569 & 0.904501 & 0.968325 & \cellcolor{green!20}$-0.000754$ & \cellcolor{green!20}$-0.000260$ & 802.0 \\
8  & gaussian\_noise\_0.15 & 0.905920 & 0.929410 & 0.904347 & 0.916707 & 0.904315 & 0.968178 & \cellcolor{green!20}$-0.000568$ & \cellcolor{green!20}$-0.000113$ & 794.7 \\
9  & gaussian\_noise\_0.20 & 0.905251 & 0.923772 & 0.909546 & 0.916603 & 0.903461 & 0.967812 & \cellcolor{red!20}0.000286 & \cellcolor{red!20}0.000253 & 788.9 \\
10 & jpeg\_quality\_90 & 0.905423 & 0.932573 & 0.899854 & 0.915921 & 0.903925 & 0.968113 & \cellcolor{green!20}$-0.000178$ & \cellcolor{green!20}$-0.000048$ & 784.2 \\
11 & jpeg\_quality\_70 & 0.905476 & 0.932839 & 0.899657 & 0.915948 & 0.903986 & 0.968192 & \cellcolor{green!20}$-0.000238$ & \cellcolor{green!20}$-0.000127$ & 799.4 \\
12 & jpeg\_quality\_50 & 0.905741 & 0.932777 & 0.900225 & 0.916212 & 0.904246 & 0.968278 & \cellcolor{green!20}$-0.000498$ & \cellcolor{green!20}$-0.000213$ & 816.5 \\
13 & jpeg\_quality\_30 & 0.906245 & 0.932611 & 0.901359 & 0.916719 & 0.904738 & 0.968350 & \cellcolor{green!20}$-0.000991$ & \cellcolor{green!20}$-0.000286$ & 862.0 \\
14 & brightness\_0.5 & 0.868409 & 0.836797 & 0.956730 & 0.892753 & 0.861260 & 0.962640 & \cellcolor{red!20}0.042488 & \cellcolor{red!20}0.005425 & 692.8 \\
15 & brightness\_0.7 & 0.904946 & 0.920034 & 0.913344 & 0.916676 & 0.903023 & 0.967931 & \cellcolor{red!20}0.000724 & \cellcolor{red!20}0.000134 & 678.9 \\
16 & contrast\_0.5 & 0.876489 & 0.849796 & 0.952631 & 0.898280 & 0.870548 & 0.964185 & \cellcolor{red!20}0.033199 & \cellcolor{red!20}0.003880 & 707.0 \\
17 & contrast\_0.7 & 0.905045 & 0.919773 & 0.913841 & 0.916798 & 0.903112 & 0.967991 & \cellcolor{red!20}0.000636 & \cellcolor{red!20}0.000074 & 702.9 \\
18 & downscale\_192$\to$224 & 0.834994 & 0.792176 & 0.964904 & 0.870050 & 0.822043 & 0.954903 & \cellcolor{red!20}0.081705 & \cellcolor{red!20}0.013162 & 844.6 \\
19 & downscale\_168$\to$224 & 0.767335 & 0.715863 & 0.984241 & 0.828869 & 0.732788 & 0.950762 & \cellcolor{red!20}0.170960 & \cellcolor{red!20}0.017303 & 832.8 \\
20 & downscale\_112$\to$224 & 0.572673 & 0.572588 & 1.000000 & 0.728211 & 0.364570 & 0.927357 & \cellcolor{red!20}0.539177 & \cellcolor{red!20}0.040708 & 813.9 \\
21 & upscale\_336$\to$224 & 0.898237 & 0.897381 & 0.928408 & 0.912631 & 0.895399 & 0.966279 & \cellcolor{red!20}0.008349 & \cellcolor{red!20}0.001786 & 944.5 \\
22 & upscale\_448$\to$224 & 0.894559 & 0.887795 & 0.933838 & 0.910235 & 0.891242 & 0.965552 & \cellcolor{red!20}0.012506 & \cellcolor{red!20}0.002513 & 1066.6 \\
23 & upscale\_672$\to$224 & 0.903441 & 0.914174 & 0.917466 & 0.915817 & 0.901308 & 0.967411 & \cellcolor{red!20}0.002439 & \cellcolor{red!20}0.000654 & 1345.2 \\
24 & occlusion\_5percent & 0.892921 & 0.914531 & 0.896763 & 0.905560 & 0.890969 & 0.959800 & \cellcolor{red!20}0.012779 & \cellcolor{red!20}0.008265 & 707.9 \\
25 & occlusion\_10percent & 0.873897 & 0.886018 & 0.894840 & 0.890407 & 0.870969 & 0.946818 & \cellcolor{red!20}0.032779 & \cellcolor{red!20}0.021247 & 713.6 \\
26 & occlusion\_15percent & 0.845447 & 0.840116 & 0.901614 & 0.869779 & 0.839856 & 0.928128 & \cellcolor{red!20}0.063892 & \cellcolor{red!20}0.039937 & 714.9 \\
27 & occlusion\_20percent & 0.809096 & 0.787016 & 0.913830 & 0.845696 & 0.797715 & 0.904108 & \cellcolor{red!20}0.106032 & \cellcolor{red!20}0.063957 & 702.3 \\

\noalign{\smallskip}\hline

\end{tabular}%
}
\endgroup
\end{table}

Therefore, QRNet showed strong robustness to Gaussian noise, JPEG
compression, changes in brightness and contrast, upscaling and small
occlusions. 
Its main weaknesses were severe blur, strong downscaling and large
occlusions, which damage or hide important parts of the QR code
structure. Under the strongest blur and downscaling conditions,
macro-F1 decreased much more than ROC--AUC. This shows that QRNet still
retained some ability to rank malicious samples above benign ones, but
its prediction scores shifted toward the malicious class at the fixed
threshold of 0.5. These results show that QRNet is reliable under common
level distortions, while further improvement is needed for heavily
blurred, low resolution, or partially covered QR codes.

\subsubsection{Explainability Results}

\begin{figure}[!htbp]
    \centering

    \makebox[\textwidth][c]{%
        \subfloat[QRNet local explainability results.
        \label{fig:qrnet-local-xai}]{%
            \includegraphics[width=0.95\textwidth]{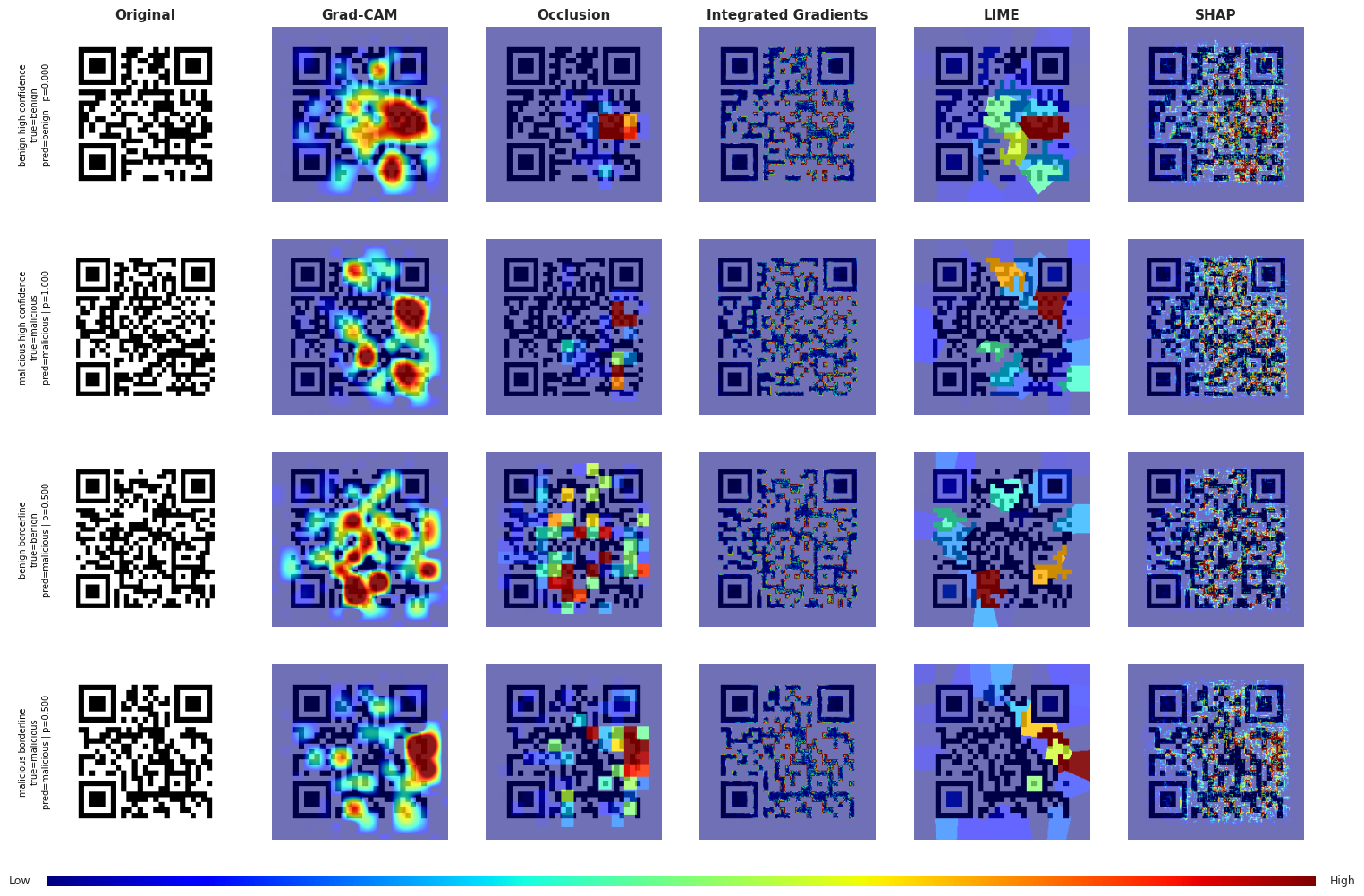}%
        }%
    }

    \par\medskip

    \makebox[\textwidth][c]{%
        \subfloat[XGBoost local explainability results.
        \label{fig:xgboost-local-xai}]{%
            \begin{minipage}{0.95\textwidth}
                \centering
                \includegraphics[width=0.49\linewidth]{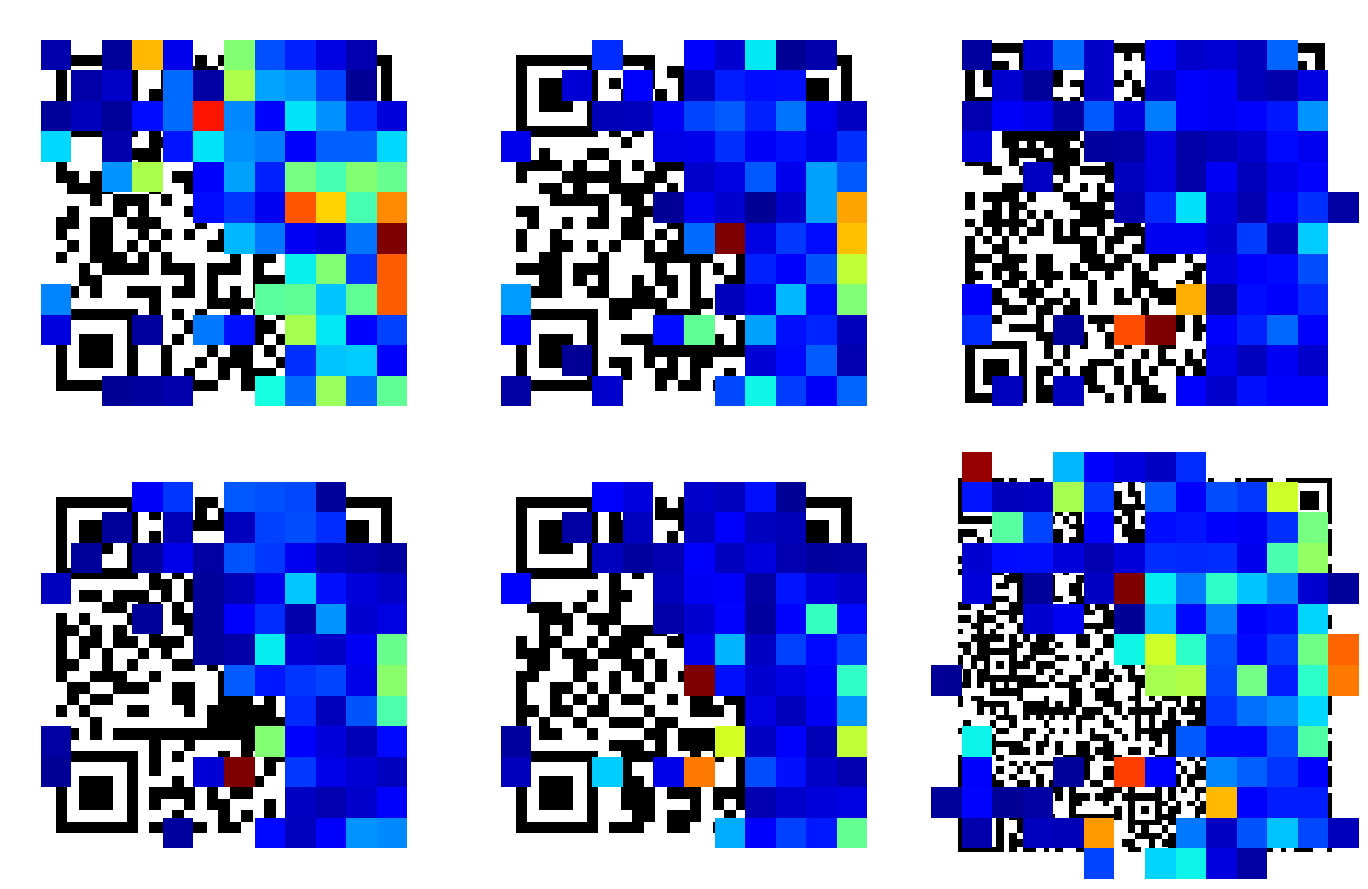}%
                \hfill
                \includegraphics[width=0.49\linewidth]{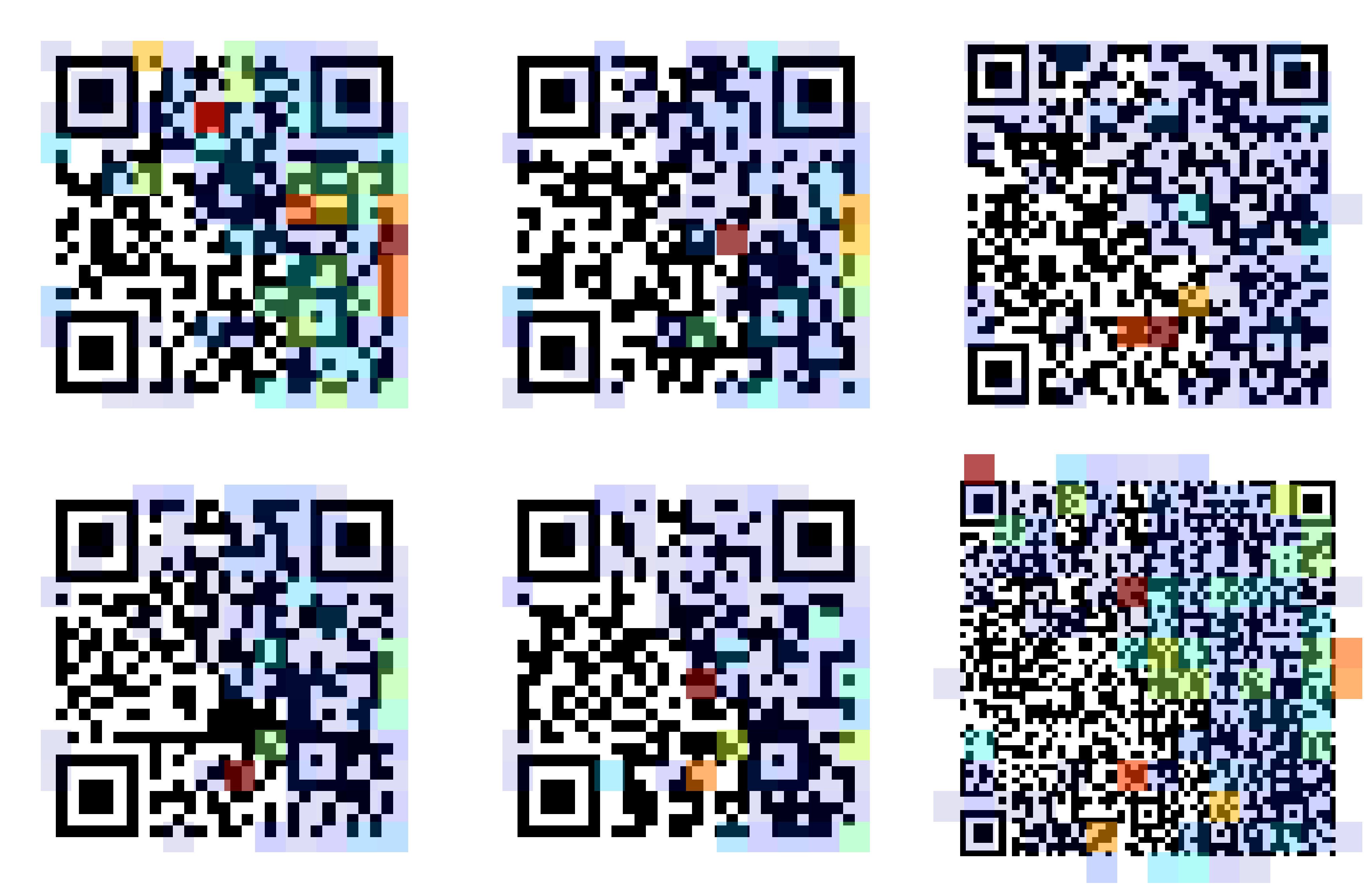}%
            \end{minipage}%
        }%
    }

    \caption{Local explainability analysis: (a) QRNet and (b) XGBoost.}
    \label{fig:local-xai-comparison}
\end{figure}

\begin{figure}[!htbp]
    \centering

    \subfloat[QRNet global SHAP beeswarm plot.
    \label{fig:qrnet-global-xai}]{%
        \begin{minipage}{0.49\textwidth}
            \centering
            \includegraphics[
                height=0.42\textheight,
                width=\linewidth,
                keepaspectratio
            ]{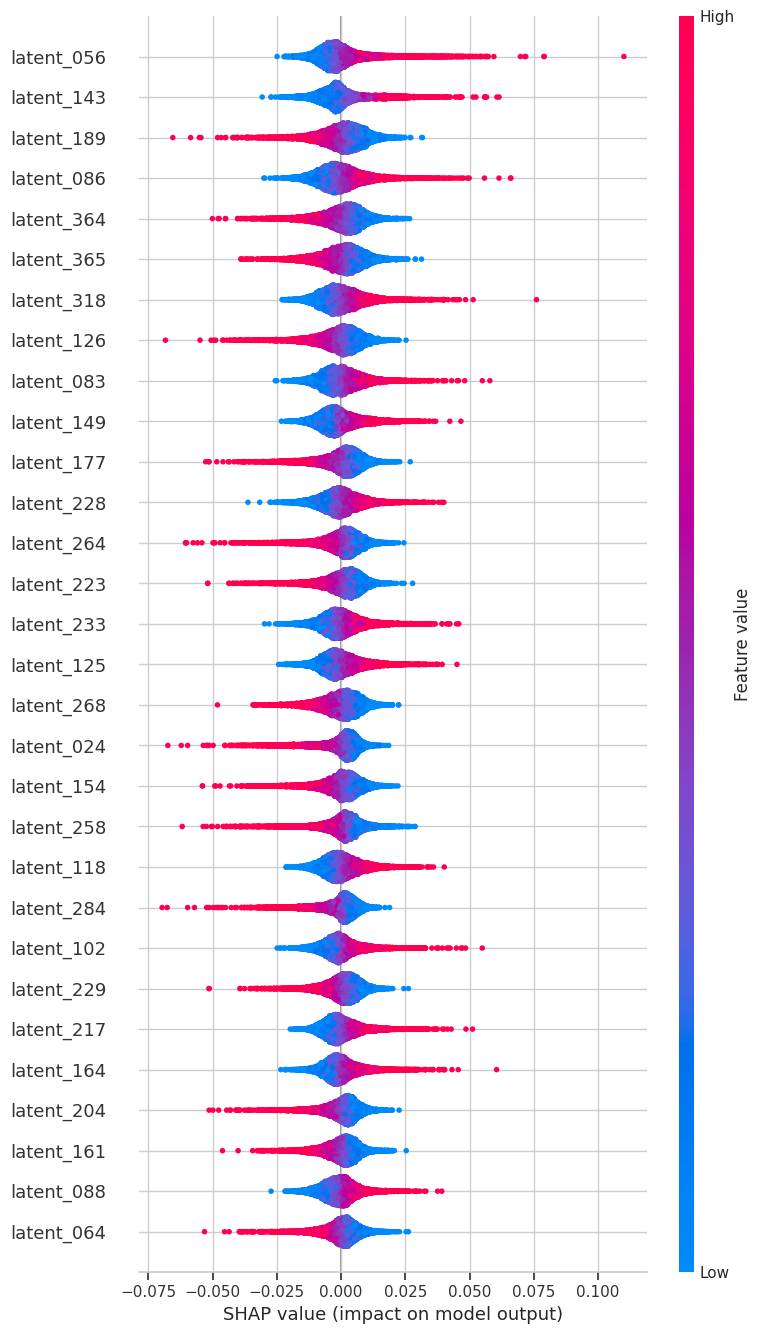}%
        \end{minipage}%
    }
    \hfill
    \subfloat[XGBoost global SHAP beeswarm plot.
    \label{fig:xgboost-global-xai}]{%
        \begin{minipage}{0.49\textwidth}
            \centering
            \includegraphics[
                height=0.42\textheight,
                width=0.85\linewidth
            ]{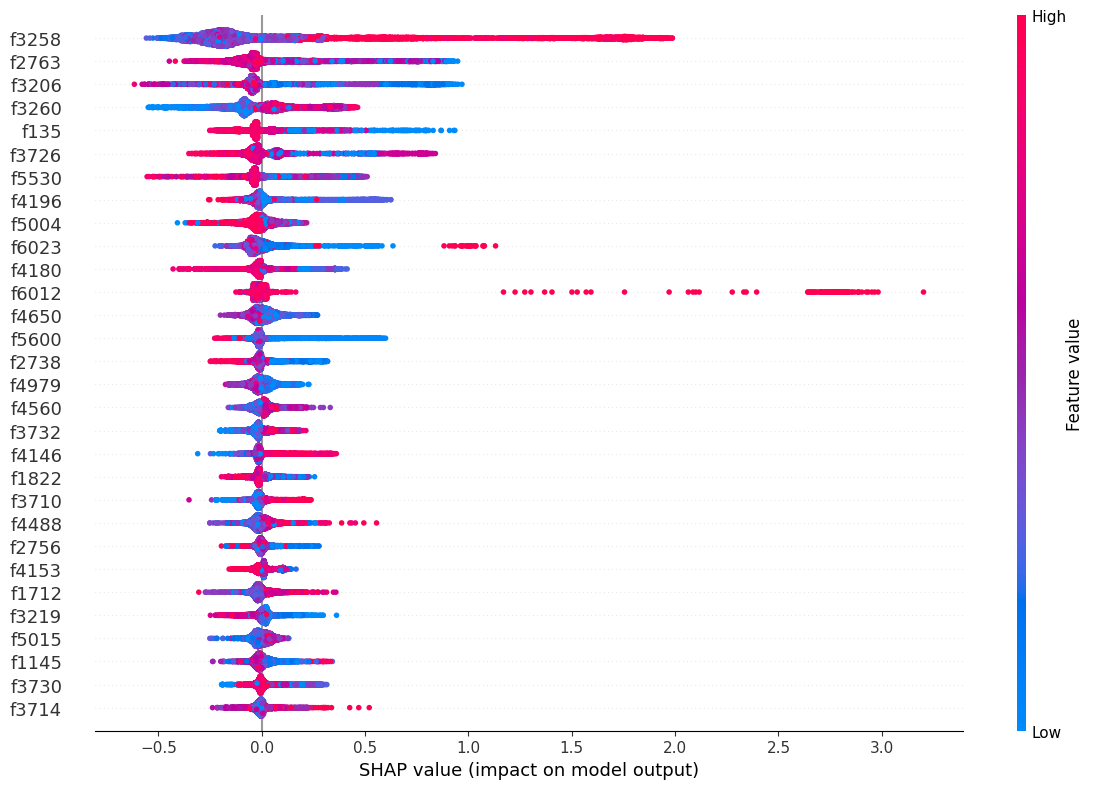}%
        \end{minipage}%
    }

    \caption{Global SHAP beeswarm analysis: (a) QRNet and (b) XGBoost.}
    \label{fig:global-xai-comparison}

\end{figure}

Explainability was applied to QRNet and XGBoost, the two best performing
models, to understand which visual regions and learned features guided
their predictions. The aim was also to examine whether both models
focused on similar parts of the QR code, even though QRNet learns its
features automatically from the QR code image, while XGBoost uses
HOG-based features. Figure \ref{fig:local-xai-comparison} presents the
local explanations, while Figure \ref{fig:global-xai-comparison} shows global SHAP explainability results across 10,000 test samples.

For QRNet, 5 local explanation methods were used, including Grad-CAM,
occlusion sensitivity, Integrated Gradients, LIME and SHAP, as shown in
Figure \ref{fig:qrnet-local-xai}. Each method explains the prediction in
a different way, but a similar visual pattern appears across them. The
stronger highlighted regions are mainly found on the right and
central right parts of the QR codes. This pattern appears across the
five methods, giving more confidence that it is not caused by only one
explanation technique. A similar result is observed for XGBoost in
Figure \ref{fig:xgboost-local-xai}. Its explanations are more
block shaped because XGBoost uses HOG features extracted from fixed
image cells, but the strongest blocks are again mainly located on the
right and central right parts of the QR codes, while the left side is
less frequently highlighted. This agreement is important because QRNet
learns its own image features, whereas XGBoost uses handcrafted HOG
features, yet both models focused on similar visual regions.

The global SHAP results show how the important features influenced the
predictions of both models. In the beeswarm plots, red points represent
high feature values and blue points represent low feature values, while
their horizontal position shows the direction of their effect. Positive
SHAP values push the prediction toward the malicious class, whereas
negative SHAP values push it toward the benign class. The QRNet plot in
Figure \ref{fig:qrnet-global-xai} shows a clear separation between these
effects. High values of \texttt{latent\_056}, \texttt{latent\_143},
\texttt{latent\_086} and \texttt{latent\_318} are mainly found on the
positive side and therefore tend to support malicious predictions, while
high values of \texttt{latent\_189}, \texttt{latent\_364},
\texttt{latent\_365} and \texttt{latent\_126} are mainly found on the
negative side and therefore support benign predictions. This makes the
QRNet plot visually clear, since the effect of high and low values can
be distinguished for several of its main latent features. In contrast,
the XGBoost plot in Figure \ref{fig:xgboost-global-xai} shows a more
concentrated pattern. Feature \texttt{f3258} has the strongest effect,
with high values mainly pushing predictions toward the malicious class.
High values of \texttt{f3260} show a similar behavior, whereas high
values of \texttt{f2763} and \texttt{f3206} mainly push predictions
toward the benign class. However, the red and blue points of several
XGBoost features are more mixed across both sides of the plot, making
their direction less visually separated than in QRNet. 
The results show that QRNet distributes importance across several
latent features, with clear positive and negative effects depending on
their values. In contrast, XGBoost shows a stronger concentration on a
smaller group of HOG features, especially \texttt{f3258}, which has the
widest SHAP spread in Figure \ref{fig:xgboost-global-xai}. This indicates
that QRNet uses a more distributed feature pattern, while XGBoost places greater importance on a smaller group of features.

The local and global results therefore show two different decision
patterns. QRNet uses several learned features together, while XGBoost
depends strongly on a smaller set of HOG features. At the same time, the local explanations of both models repeatedly highlight the right and central right parts of the QR codes, while the left side is rarely highlighted. These areas contain both QR structural elements and encoded data modules, suggesting that patterns in this region provided useful information for separating benign and malicious samples. This common focus is important because QRNet and XGBoost use very different feature representations, yet both models identify similar areas of the QR code as important for prediction.

\subsection{Generalization and Transfer Evaluation}

\subsubsection{QR Specific Transfer Tasks}

\begin{figure}[!htbp]
    \centering
    \includegraphics[width=0.65\textwidth]{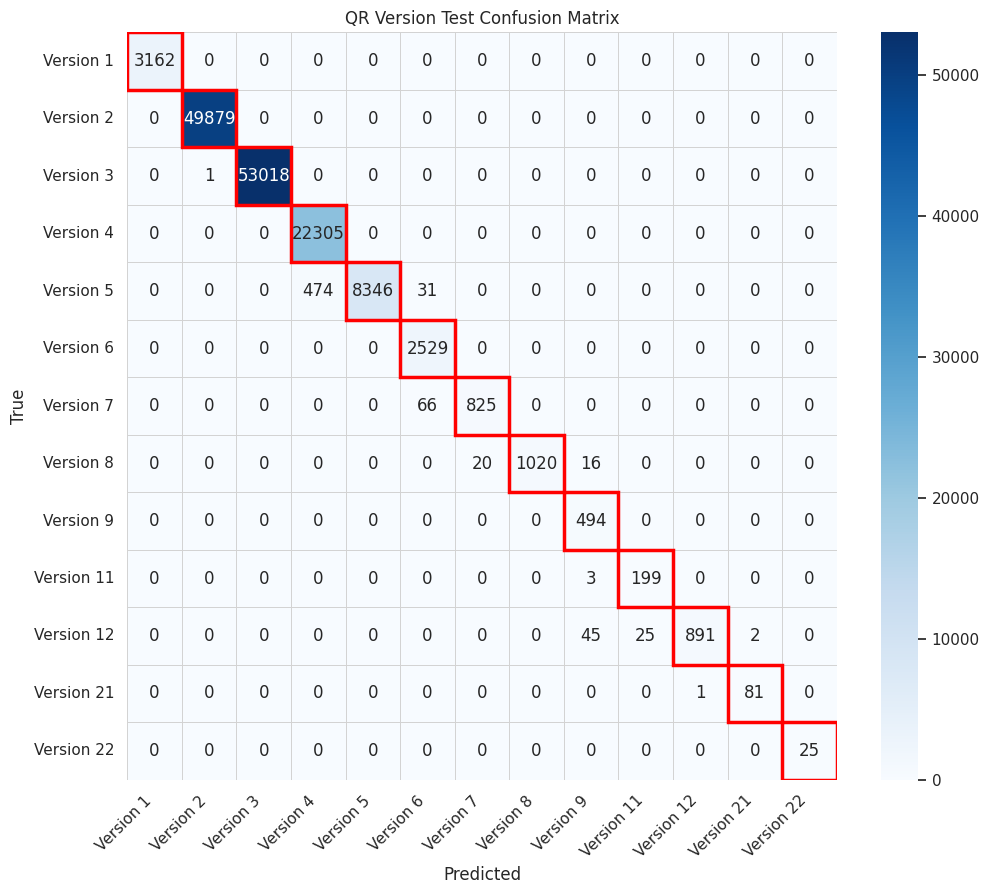}
    \caption{QR version confusion matrix.}
    \label{fig:qr-version-cm}
\end{figure}

\begin{figure}[!htbp]
    \centering
    \includegraphics[width=0.65\textwidth]{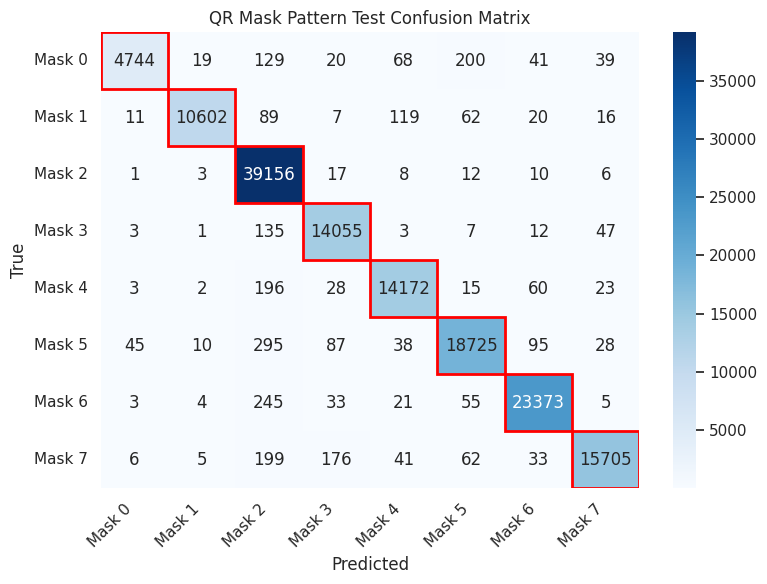}
    \caption{QR mask pattern confusion matrix.}
    \label{fig:qr-mask-cm}
\end{figure}

The transferability of the QRNet representation was evaluated on two
QR specific structural tasks, QR version classification and mask pattern
classification. These experiments examine whether the features learned
for benign and malicious QR code classification also preserve information
about the internal structure of QR codes. Both transfer tasks were
derived from the complete 1,005,738 image dataset, with only images for
which valid structural labels could be extracted for training
and evaluation. Table \ref{tab:transfer-external-performance}
summarizes the overall results, while
Figures \ref{fig:qr-version-cm} and \ref{fig:qr-mask-cm} show the
corresponding confusion matrices.

\textbf{For QR version classification}, valid version labels were obtained for 956,411 images and 956,386 images from 13 QR versions were used after
removing 25 images from rare unsupported versions. On the test set of
143,458 images, QRNet achieved an accuracy of 0.9952, a macro-precision
of 0.9743, a macro-recall of 0.9794, a macro-F1 of 0.9761 and a
ROC-AUC of 0.9995, as reported in Table \ref{tab:transfer-external-performance}. The top 3 accuracy was
0.999979, showing that the correct version was almost always among the
three highest ranked predictions. The confusion matrix in
Figure \ref{fig:qr-version-cm} shows that most predictions fall on the
main diagonal, with only 684 test images misclassified out of 143,458.
The largest confusion occurred between Versions 5 and 4, where 474
Version 5 images were predicted as Version 4. Smaller errors were also
seen for Versions 7, 9, 11 and 12. Detailed class wise precision, recall, F1-score and ROC--AUC results are provided in Table \ref{tab:qr-version-classification-report} in the Appendix.

\textbf{For mask pattern classification}, valid labels were obtained for
956,330 images covering all eight QR mask patterns. On the test set of
143,450 images, QRNet achieved an accuracy of 0.9796, a macro-precision
of 0.9824, a macro-recall of 0.9693, a macro-F1 of 0.9756 and a
ROC-AUC of 0.9953 as reported in
Table \ref{tab:transfer-external-performance}. The top 3 accuracy was
0.9929, showing that the correct mask was usually among the three
highest ranked predictions. The confusion matrix in
Figure \ref{fig:qr-mask-cm} shows that most predictions fall on the
main diagonal, with 2,918 test images misclassified out of 143,450.
Mask 0 was the most difficult class because it had the lowest recall,
while Mask 6 achieved the highest F1-score. Many of the errors involved
predictions as Mask 2, including 295 Mask 5 images, 245 Mask 6 images,
and 199 Mask 7 images. Detailed class wise precision, recall, F1-score, and ROC--AUC results are provided in Table \ref{tab:qr-mask-classification-report} in the Appendix.

The two tasks had different levels of difficulty. QR version
classification achieved higher accuracy because the QR version changes
the size, module width and visual density of the QR code, making the
versions easier to distinguish. Mask pattern classification was more
difficult because the eight masks keep the same overall QR structure
and mainly differ in how the data modules are arranged. This means the
model had to learn smaller visual differences between the mask patterns.
Despite this difference, both tasks achieved very high and similar
results across the 5 reported metrics. These results show that the
QRNet representation is not limited to benign/malicious classification
and can also capture useful structural information for distinguishing
QR versions, mask patterns and potentially other QR specific tasks.

\subsubsection{External Dataset Evaluation}

\begin{figure}[!htbp]
    \centering

    \subfloat[Zero-shot test set evaluation.
    \label{fig:external-zero-shot-test}]{%
        \includegraphics[width=0.32\textwidth]{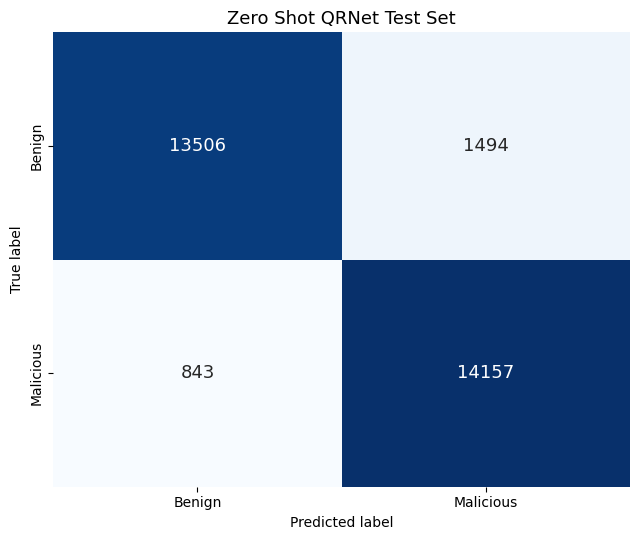}%
    }
    \hfill
    \subfloat[Fine-tuned test set evaluation.
    \label{fig:external-finetuned-test}]{%
        \includegraphics[width=0.32\textwidth]{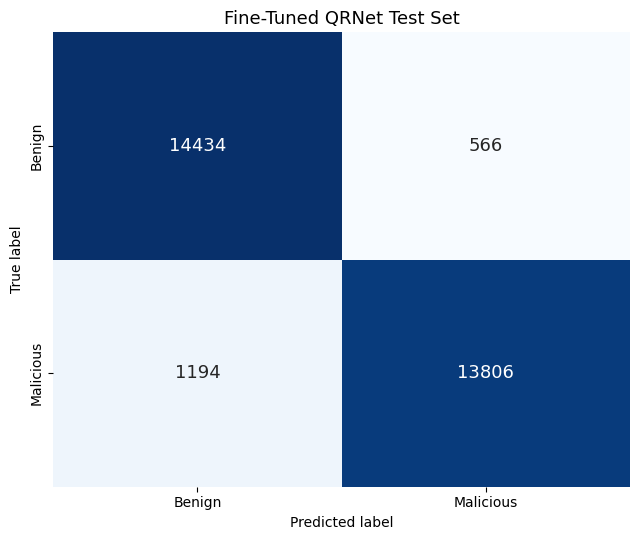}%
    }
    \hfill
    \subfloat[Zero-shot full dataset evaluation.
    \label{fig:external-zero-shot-full}]{%
        \includegraphics[width=0.32\textwidth]{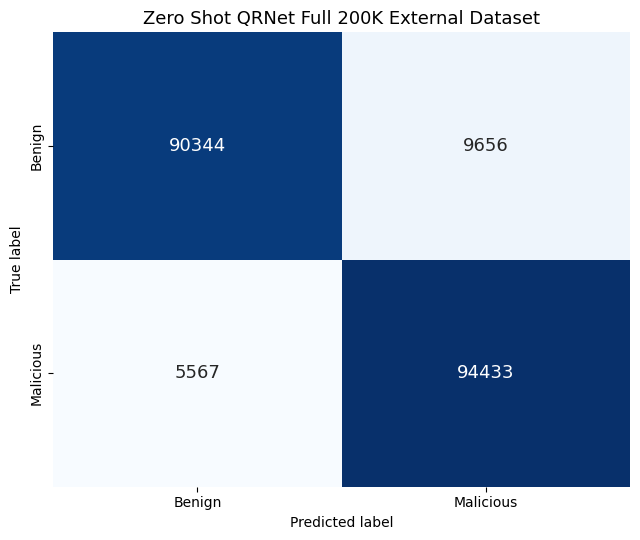}%
    }

    \caption{QRNet confusion matrices on the external QR code dataset
    under different evaluation scenarios.}
    \label{fig:external-dataset-confusion-matrices}

\end{figure}

QRNet was evaluated on an independent external QR code image dataset,
containing 200,000 images with equal numbers of benign and malicious
images. This external dataset is separate from the dataset used to develop QRNet and was used to test how well the model performs on QR code data from a different dataset and whether its learned features can generalize beyond the original data. Table \ref{tab:transfer-external-performance} summarizes the main results, while Figure \ref{fig:external-dataset-confusion-matrices} shows the confusion matrices for the different evaluations.

\begin{table}[!htbp]
\centering
\caption{Performance of QRNet on QR Structural Classification and External Dataset Evaluation.}
\label{tab:transfer-external-performance}

\begingroup
\normalsize
\setlength{\tabcolsep}{4pt}
\renewcommand{\arraystretch}{1.15}
\sisetup{detect-weight=true}

\begin{tabularx}{\textwidth}{@{}
>{\raggedright\arraybackslash}X
*{5}{S[table-format=1.4]}@{}}

\hline\noalign{\smallskip}
\textbf{Task} &
{\textbf{Accuracy}} &
{\shortstack{\textbf{Macro-}\\\textbf{Precision}}} &
{\shortstack{\textbf{Macro-}\\\textbf{Recall}}} &
{\textbf{Macro-F1}} &
{\textbf{ROC-AUC}} \\

\noalign{\smallskip}\hline\noalign{\smallskip}

QR Mask Pattern Classification (1 million) &
0.9796 & 0.9824 & 0.9693 & 0.9756 & 0.9953 \\

QR Version Classification (1 million) &
0.9952 & 0.9743 & 0.9794 & 0.9761 & 0.9995 \\

Zero-Shot Quishing Detection on New Dataset
\cite{ahsan2026qrcodequishing200k} &
0.9238 & 0.9245 & 0.9238 & 0.9238 & 0.9803 \\

Fine-tuned Quishing Detection on New Dataset
\cite{ahsan2026qrcodequishing200k} &
0.9413 & 0.9421 & 0.9413 & 0.9413 & 0.9865 \\

\noalign{\smallskip}\hline

\end{tabularx}
\endgroup
\end{table}

On the external dataset \cite{ahsan2026qrcodequishing200k}, QRNet
was first tested without any additional training. On the full
200,000 images dataset, it achieved an accuracy of 0.9238, a
macro-precision of 0.9245, a macro-recall of 0.9238, a macro-F1 of
0.9238 and a ROC-AUC of 0.9803. As shown in Figure \ref{fig:external-zero-shot-full}, 15,223 out of 200,000 images
were misclassified. The zero-shot model was also tested on the fixed 30,000 image test split, where 2,337 images were misclassified. After fine-tuning on the training split of the external dataset and
evaluating on the same fixed test split, all five reported metrics improved. Accuracy increased to 0.9413, while macro-precision, macro-recall, macro-F1 and ROC-AUC reached 0.9421,
0.9413, 0.9413 and 0.9865, respectively.
Figure \ref{fig:external-finetuned-test} shows that the number of errors
on the same test split decreased from 2,337 to 1,760, while false
positives decreased from 1,494 to 566.

The external evaluation shows that QRNet can maintain strong performance
on unseen QR code image data without retraining and can improve further after fine-tuning on labeled data from a new dataset. Together with the
QR specific transfer results, these findings show that QRNet is not
limited to benign/malicious classification. Its learned representation
can also be reused for other QR related tasks, such as QR version and
mask pattern classification, and it can adapt to data from a different QR code
dataset. QRNet can therefore be used as a QR specific feature extractor
or backbone, similar to general image models such as DenseNet, ResNet,
and EfficientNet, while being much smaller in size and used specifically
for QR code images. Combined with the earlier numerical and statistical comparisons, robustness evaluation and explainability results, these findings support the use of QRNet as a strong and reusable model for a range of QR specific applications.

\section{Conclusion and Future Work}
\label{sec:conclusion}

\subsection{Conclusion}

This work introduced QRNet, a novel deep learning architecture for
QR code image analysis and quishing attack detection. QRNet learns directly
from the visual structure of QR code images without requiring URL
decoding or manual feature extraction. On the full test set, the
model achieved an accuracy of 0.9053, an F1-score of 0.9158 and a ROC-AUC of 0.9681. QRNet also achieved better overall performance than all the evaluated machine learning and deep learning baseline models, with the statistical tests supporting the observed differences. These results show that visual information in QR codes can provide useful features for detecting benign and malicious samples.

The additional experiments showed that the QRNet representation is not
limited to the original benign/malicious classification task. The model
remained stable under several image distortions, while the explainability
analysis identified the QR code regions and learned features that had the
largest effect on its predictions. QRNet also transferred successfully to
QR version and mask pattern classification, with macro-F1 scores of
0.9761 and 0.9756, respectively, together with near perfect ROC-AUC
values of 0.9995 and 0.9953. In addition, QRNet showed strong performance on the independent external QR code dataset. Zero-shot accuracy reached 0.9238 with a ROC-AUC of 0.9803, improving to 0.9413 and 0.9865, respectively, after fine-tuning.

These findings show that QRNet can be used for more than quishing
attack detection. Its strong classification performance, statistical
validation, robustness, explainability, transfer results, and external
generalization support its use as a QR specific feature extractor or
backbone for different QR code applications. By learning directly from
QR code images and reusing the same representation across different
tasks and dataset, QRNet provides a focused and better alternative to general image architectures for QR related problems. These results support QRNet as a strong and reusable deep learning architecture for QR code security and other QR specific applications.

\subsection{Future Work}

Future work can extend QRNet to real world QR code applications.
The model could be integrated into mobile phones, QR scanners, browser
tools or security applications to check a QR code before the user opens
its content. This would allow QRNet to be studied in practical settings
where QR codes are captured using different cameras, distances, angles,
lighting conditions, and from printed copies. Methods such as continual
learning, self-supervised learning, domain adaptation and test time
adaptation could also help the model handle new QR code patterns and
changing quishing campaigns over time. In addition, quantization,
pruning and knowledge distillation could be explored to further reduce
the model size and inference time for mobile and edge deployment.

Another direction is to study QRNet against adversarial QR code attacks.
These may include small module changes, added logos, overlays, stickers
or other changes made to mislead the classifier while keeping the QR code readable. Such experiments could help study how QRNet behaves when an attacker intentionally changes the visual structure of the QR code and could lead to stronger defenses against these attacks. Additionally, future work could evaluate QRNet on colored and stylized QR codes used in social media, payment systems and restaurant menus to assess its generalization to diverse real world visual designs.

Future work could also explore the use of vision language models (VLMs), large language models (LLMs) and other recent foundation models for QR code security. Their performance could be compared with QRNet in terms of detection metrics, model size, inference time, and computational cost. It would also be useful to examine whether these larger models can learn QR specific patterns as effectively as a smaller model trained directly on QR code images.

\appendix

\renewcommand{\thesection}{Appendix \Alph{section}}
\section{QR Specific Structural Classification Results}

\begin{table}[H]
\centering
\caption{Classification report for QR version classification.}
\label{tab:qr-version-classification-report}

\begingroup
\normalsize
\setlength{\tabcolsep}{6pt}
\renewcommand{\arraystretch}{1.15}

\begin{tabular}{@{}lrrrrr@{}}

\hline\noalign{\smallskip}
\textbf{Class} & \textbf{Precision} & \textbf{Recall} &
\textbf{F1-score} & \textbf{ROC--AUC} &
\textbf{Support} \\
\noalign{\smallskip}\hline\noalign{\smallskip}

Version 1  & 1.000000 & 1.000000 & 1.000000 & 1.000000 & 3162 \\
Version 2  & 0.999980 & 1.000000 & 0.999990 & 0.999999 & 49879 \\
Version 3  & 1.000000 & 0.999981 & 0.999991 & 0.999994 & 53019 \\
Version 4  & 0.979191 & 1.000000 & 0.989486 & 0.998657 & 22305 \\
Version 5  & 1.000000 & 0.942944 & 0.970634 & 0.996389 & 8851 \\
Version 6  & 0.963062 & 1.000000 & 0.981183 & 0.999748 & 2529 \\
Version 7  & 0.976331 & 0.925926 & 0.950461 & 0.999497 & 891 \\
Version 8  & 1.000000 & 0.965909 & 0.982659 & 0.999859 & 1056 \\
Version 9  & 0.885305 & 1.000000 & 0.939163 & 0.999834 & 494 \\
Version 11 & 0.888393 & 0.985149 & 0.934272 & 0.999902 & 202 \\
Version 12 & 0.998879 & 0.925234 & 0.960647 & 0.999859 & 963 \\
Version 21 & 0.975904 & 0.987805 & 0.981818 & 0.999999 & 82 \\
Version 22 & 1.000000 & 1.000000 & 1.000000 & 1.000000 & 25 \\

\noalign{\smallskip}\hline\noalign{\smallskip}

Accuracy     & 0.995232 & 0.995232 & 0.995232 & 0.995232 & 143458 \\
Macro avg    & 0.974388 & 0.979457 & 0.976177 & 0.999518 & 143458 \\
Weighted avg & 0.995386 & 0.995232 & 0.995203 & 0.999555 & 143458 \\

\noalign{\smallskip}\hline

\end{tabular}
\endgroup
\end{table}

\begin{table}[H]
\centering
\caption{Classification report for QR mask pattern classification.}
\label{tab:qr-mask-classification-report}

\begingroup
\normalsize
\setlength{\tabcolsep}{6pt}
\renewcommand{\arraystretch}{1.15}

\begin{tabular}{@{}lrrrrr@{}}

\hline\noalign{\smallskip}
\textbf{Class} & \textbf{Precision} & \textbf{Recall} &
\textbf{F1-score} & \textbf{ROC--AUC} &
\textbf{Support} \\
\noalign{\smallskip}\hline\noalign{\smallskip}

Mask 0 & 0.985050 & 0.901901 & 0.941644 & 0.988286 & 5260 \\
Mask 1 & 0.995867 & 0.970346 & 0.982941 & 0.996976 & 10926 \\
Mask 2 & 0.968153 & 0.998546 & 0.983115 & 0.996636 & 39213 \\
Mask 3 & 0.974485 & 0.985417 & 0.979921 & 0.998263 & 14263 \\
Mask 4 & 0.979406 & 0.977447 & 0.978425 & 0.996185 & 14499 \\
Mask 5 & 0.978420 & 0.969052 & 0.973714 & 0.994680 & 19323 \\
Mask 6 & 0.988538 & 0.984582 & 0.986556 & 0.997058 & 23739 \\
Mask 7 & 0.989665 & 0.967831 & 0.978627 & 0.994899 & 16227 \\

\noalign{\smallskip}\hline\noalign{\smallskip}

Accuracy     & 0.979658 & 0.979658 & 0.979658 & 0.979658 & 143450 \\
Macro avg    & 0.982448 & 0.969390 & 0.975618 & 0.995373 & 143450 \\
Weighted avg & 0.979840 & 0.979658 & 0.979585 & 0.996082 & 143450 \\

\noalign{\smallskip}\hline

\end{tabular}
\endgroup
\end{table}

\bibliographystyle{unsrtnat}
\bibliography{references}

\end{document}